\documentclass[preprint,12pt]{elsarticle}

\usepackage{amssymb}
\usepackage{svg}
\usepackage{float}
\usepackage{subfig}
\usepackage[hyphens]{url}
\usepackage{bibunits}
\usepackage{mdframed}

\usepackage{caption}

\mdfsetup{%
    linewidth=0pt,
    innerleftmargin=0pt,
    innerrightmargin=0pt,
    innertopmargin=0pt,
    innerbottommargin=0pt
}

\defaultbibliography{references}

\usepackage{comment}

\journal{arXiv}

\begin{document}

\begin{frontmatter}



\title{Power sector effects of alternative options for de-fossilizing heavy-duty vehicles -- go electric, and charge smartly}

\author[DIW]{Carlos Gaete-Morales}
\author[ifeu]{Julius Jöhrens}
\author[ifeu]{Florian Heining}
\author[DIW]{Wolf-Peter Schill\corref{cor}}

\cortext[cor]{Corresponding author \& lead contact \ead{wschill@diw.de}}

\affiliation[DIW]{organization={German Institute for Economic Research (DIW Berlin)},
            addressline={Mohrenstraße 58}, 
            city={Berlin},
            postcode={10117}, 
            country={Germany}}

\affiliation[ifeu]{organization={ifeu - Institute for Energy and Environmental Research},
            addressline={Wilckensstraße 3}, 
            city={Heidelberg},
            postcode={69120}, 
            country={Germany}}

\begin{abstract}
Various options are discussed to de-fossilize heavy-duty vehicles (HDV), including battery-electric vehicles (BEV), electric road systems (ERS), and indirect electrification via hydrogen fuel cells or e-fuels. We investigate their power sector implications in future scenarios of Germany with high renewable energy shares, using an open-source capacity expansion model and route-based truck traffic data. Power sector costs are lowest for flexibly charged BEV that also carry out vehicle-to-grid operations, and highest for e-fuels. If BEV and ERS-BEV are not optimally charged, power sector costs increase, but are still substantially lower than in scenarios with hydrogen or e-fuels. This is because indirect electrification is less energy efficient, which outweighs potential flexibility benefits. BEV and ERS-BEV favor solar photovoltaic energy, while hydrogen and e-fuels favor wind power and increase fossil electricity generation. Results remain qualitatively robust in sensitivity analyses.

\end{abstract}




\end{frontmatter}


\begin{bibunit}

\section{Introduction}\label{sec: introduction}
Making energy consumption climate neutral in all end-use sectors is of paramount importance for mitigating climate change \cite{de_coninck_strengthening_2018}. A key strategy for achieving this is to substitute fossil fuels by renewable electricity, facilitated by direct or indirect electrification of end uses in mobility, heating, and industrial applications \cite{shukla_contribution_2022}. In the transportation sector, battery-electric vehicles (BEV) have emerged as the most promising option for de-fossilizing the passenger car segment. Already today, BEV can lead to sizeable greenhouse gas emission reductions compared to internal combustion engines \cite{hoekstra_underestimated_2019}, which will further increase when the electricity mix becomes cleaner. In many countries, markets for electric passenger cars have been soaring in the past years, and are expected to continue to grow strongly in the near future \cite{iea_global_2023}. For heavy-duty vehicles (HDV), however, the technology space still appears to be more open. While the feasibility of pure battery-electric HDV has been assessed to be increasing \cite{nykvist_feasibility_2021}, they compete with other options. This includes electric road systems (ERS), which allow for dynamic power transfer to electric vehicles on the road \cite{boltze_insights_2020, speth_comparing_2021}; trucks with hydrogen fuel cells; or conventional HDV with internal combustion engines that use power-to-liquid (PtL), also referred to as e-fuels, which are produced mostly with renewable electricity \cite{hannula_nearterm_2019, lajevardi_simulating_2022, plotz_hydrogen_2022,li_transition_2022}. 

These options of direct or indirect electrification of HDV have different properties concerning, on the one hand, energy efficiency, and, on the other hand, temporal flexibility of electricity use. For example, direct electrification via BEV is more energy efficient compared to indirect electrification via electrolysis-based hydrogen or e-fuels \cite{ueckerdt_2021,lajevardi_simulating_2022}. Yet, the temporal flexibility of BEV may be constrained by charging availability and limited battery capacities, as vehicle batteries are costly and heavy. In contrast, indirect electrification via hydrogen or e-fuels may entail large-scale and low-cost storage options \cite{taljegard_2017, stoeckl_2021}, but the overall energy efficiency of these supply chains is lower compared to BEV. Temporal power sector flexibility becomes increasingly important with growing shares of renewables, as the potential for firm renewable generation such as hydropower, bioenergy, or geothermal power is limited in many countries. In contrast, wind and solar power potentials are often abundant, but they have variable generation profiles that depend on weather conditions and daily and seasonal cycles \cite{lopezprol_economics_2021}. Integrating growing shares of variable renewable energy sources thus requires an increasing use of temporal and spatial flexibility options in the power sector. Flexibility may be provided on the supply side (e.g.,~dispatchable generators) or the demand side (e.g.,~flexible loads), and via energy storage or transmission \cite{kondziella_flexibility_2016,schill_joule_2020}.

\begin{figure}[ht!]
    \begin{mdframed}
    \centering
        \includegraphics[width =0.98\textwidth]{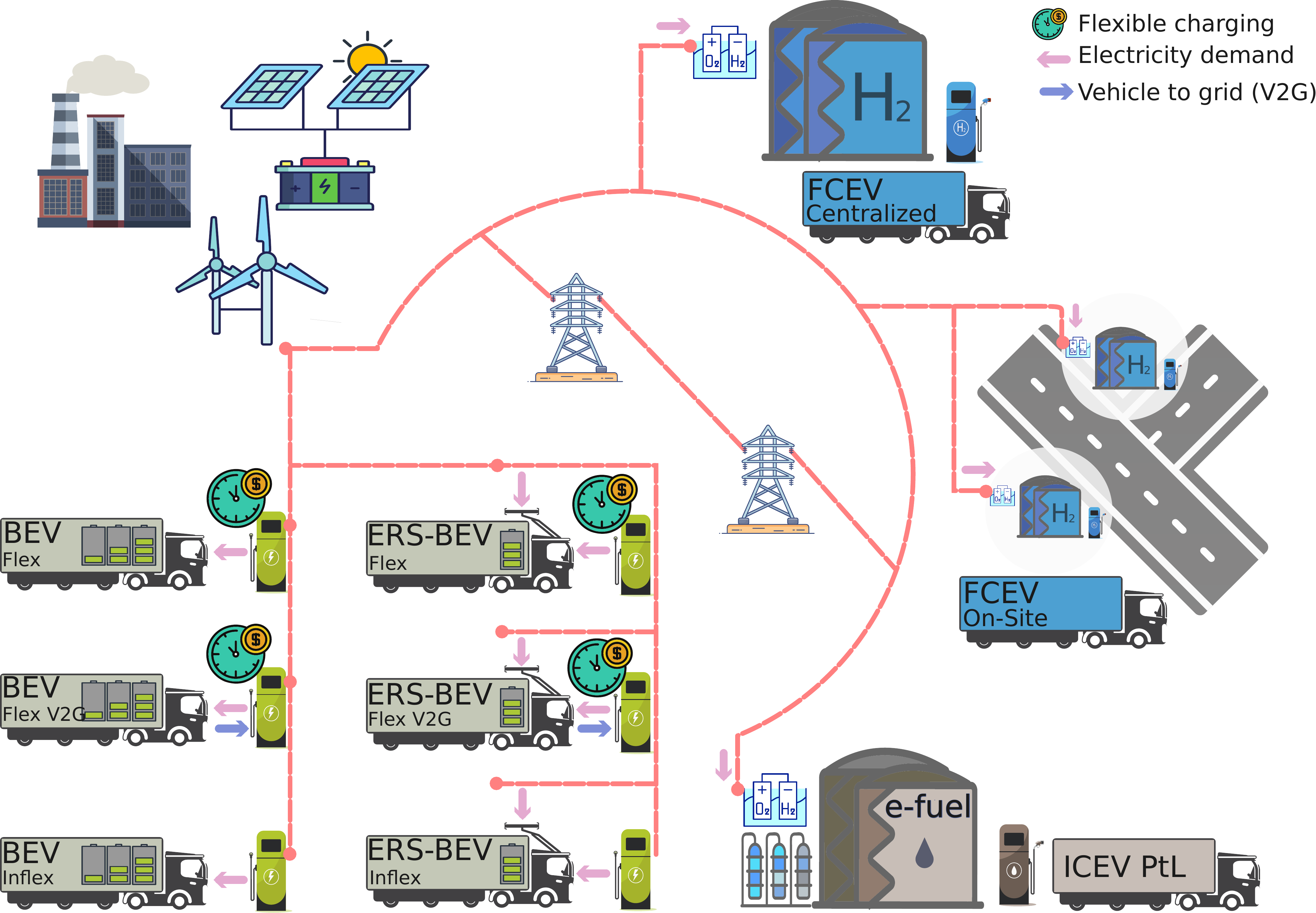}
    \end{mdframed}
    \caption{Overview of direct and indirect HDV electrification options covered in this analysis. For direct electrification via BEV and ERS-BEV we differentiate scenarios with flexible charging without and with V2G, as well as scenarios with inflexible, i.e.,~non-optimized charging. Hydrogen and PtL options are assumed to be always operated flexibly, constrained by hydrogen or e-fuel storage capacities, and do not allow reconversion to electricity.}
    \label{fig_01}
\end{figure}

Against this background, we investigate the power sector implications of different options for (in-)directly electrifying HDV, particularly focusing on the trade-off between energy efficiency and temporal flexibility. To do so, we apply an open-source capacity expansion model \cite{zerrahn_long-run_2017,gaete_2021} to scenarios of the Central European power sector for the year 2030 with high renewable energy shares. We focus on the domestic traffic of HDV in Germany with a gross vehicle weight above 26 tonnes, drawing on a detailed data set of truck trips on inner-German routes. We include stationary-charged BEV trucks as well as hybrid battery-catenary trucks as a particular example of an electric road system technology (ERS-BEV)\footnote{In general, there are three options available for ERS: Overhead catenary lines, conductor rails, and inductive power transfer. We have opted for the first option as it is the technology currently under discussion in Germany with three test tracks in place \cite{Hacker2023}.}, fuel-cell hydrogen electric trucks (FCEV), and such with internal diesel combustion engines powered by e-fuels (ICEV PtL), see Figure~\ref{fig_01}. For hydrogen and e-fuels, we assume domestic supply chains. We compare the power sector costs\footnote{Power sector costs comprise fixed and variable costs of all electricity generation and storage technologies over a full year. They do not include the costs of HDV electrification infrastructure, e.g.,~overhead lines, charging stations, or filling stations for hydrogen or PtL.} of these options, as well as their repercussions on the optimal power plant fleet and its use.

While there is a broad literature on the potential power sector impacts of battery-electric passenger cars \cite{richardson_electric_2013,muratori_shape_2020,mangipinto_impact_2022}, according research for electric HDV is sparse. A common finding of analyses focusing on passenger cars is that controlled charging can help to avoid problematic load peaks, to integrate renewable electricity generation, and to lower overall system costs \cite{schill_power_2015,gnann_2018,sadeghian_a_2022}. In particular, feeding back electricity from car batteries to the system (vehicle-to-grid, V2G) can help to balance the diurnal variability of solar photovoltaic (PV) generation \cite{brown_2018}.

How electrified HDV interact with the power sector is hardly explored so far. Some analyses investigate grid or peak load impacts without using power sector models. For example, an analysis of the distribution grid impacts of BEV-HDV with depot charging finds that many existing substations in the US would allow for massive HDV charging without grid upgrades \cite{borlaug_heavyduty_2021}. Earlier studies conclude that local grid impacts of electrified HDV near logistics centres and along major roads may be substantial in Switzerland and Finland \cite{liimatainen_2019}, or that ERS-HDV could strongly increase the peak load in Norway \cite{taljegard_2017}. Using capacity expansion power sector models, other analyses come to deviating conclusions on the impacts of electric HDV on the overall load profile and the wider power sector; in a Scandinavian-German case study that includes ERS-HDV in a subset of scenarios, effects are pronounced \cite{taljegard_impacts_2019}, while in a European model study the impacts are moderate \cite{plotz_impact_2019}. Importantly, both analyses assume exogenous, i.e.,~inflexible, charging profiles of HDV and do not temporally optimize their (dis-)charging operations. Given this research gap, we deliberately focus our analysis on the power sector effects of electrified HDV, modelling in detail their repercussions on optimal capacities and their hourly use. In doing so, we complement other strand of the literature which have focused on the total costs of ownership (TCO) \cite{nykvist_feasibility_2021,hunter_spatial_2021,samet_2024}, energy efficiency\cite{noll_analyzing_2022}, or potential health and climate impacts of electric HDV \cite{lin_2021}.

We contribute to the literature with a dedicated analysis of the power sector impacts of different HDV electrification options that co-optimizes their charging and discharging operations (including V2G) with the capacity and dispatch decisions in the power sector. We do so for a wide range of HDV technologies, including dynamic power supply via electric road systems. We use a power sector model that fully captures the hourly variability of load and renewable generation over all hours of a full year, and apply it to a future scenario with high shares of variable renewables. The model code and all input data, including detailed hourly HDV mission profiles for domestic transport in Germany, are provided open-source for transparency and reproducibility (see section~\ref{sec: exp proc}).

\section{Results}\label{sec: results}
In the main part of this paper, we show results for a setting where Germany is interconnected with its neighbors, defined as the baseline model specification. Each country is modeled as a single node. While this configuration can be considered to be a policy-relevant scenario for 2030, where electric HDV in Germany may benefit from the flexibility provided by the European interconnection, we also provide results for a sensitivity analysis where the German power sector is modeled in isolation in the Supplemental Information (\ref{sec: sensitivity island}).\footnote{In the SI, we also show further results for baseline assumptions (sections~\ref{sec: si storage}-\ref{sec: si time series}), a sensitivity analysis that explores the effects of an upper bound for wind power capacity (section~\ref{sec: sensitivity wind}), and another sensitivity with alternative charging availability assumptions (section~\ref{sec: sensitivity charging availability}).} 

We compare nine HDV scenarios against a reference without any electrification of HDV (Ref). Three scenarios cover pure BEV trucks with varying degrees of flexibility: these are either charged as flexibly as possible (BEV~Flex), additionally have the option of feeding back electricity to the grid (BEV~Flex~V2G), or are charged in a way that is not co-optimized with the power sector (BEV~Inflex). There are three respective scenarios for ERS-BEV, which can also draw electricity from overhead lines and have smaller batteries than pure BEV. Two scenarios cover fuel-cell trucks with either centralized electrolysis and transport via liquefied hydrogen, which enables low-cost hydrogen storage (FCEV~Centralized), or temporally inflexible on-site electrolysis at filling stations (FCEV~On-Site). The last scenario represents conventional trucks with internal combustion engines powered by e-fuels (ICEV PtL). To separate effects, we assume that the whole domestic fleet of HDV (318,700 vehicles~\textgreater{}~26 t) in the year 2030 consists of the particular vehicle type (compare~\ref{fig_01}). While this is not a policy-relevant assumption for 2030, our aim is to illustrate corner solutions. The power sector effects of mixed real-world HDV fleets are likely between the ones of the extreme cases considered here.

\subsection{Lowest power sector costs and electricity prices for BEV with V2G}

\begin{figure}[ht!]
    \begin{mdframed}
        \centering
        \includegraphics[width =0.75\textwidth]{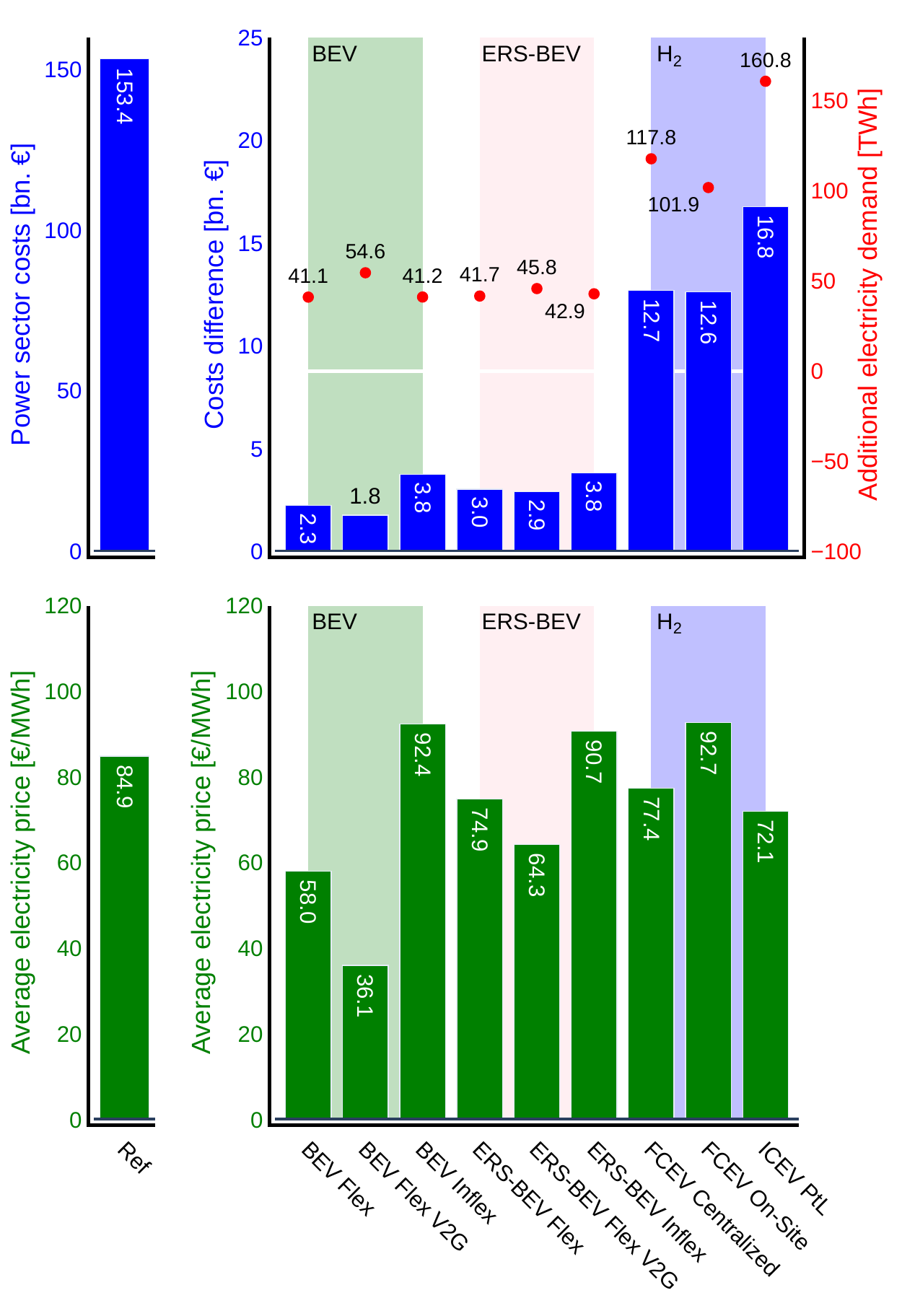}
    \end{mdframed}
    \caption{Changes in yearly power sector costs and electricity demand induced by different HDV options (upper panel), and average wholesale market prices of charging electricity (lower panel).}
    \label{fig_02}
\end{figure}

Compared to the reference case without electrified trucks, yearly power sector costs increase in all scenarios with electrified HDV (Figure~\ref{fig_02}, upper panel). The cost of the additional electricity demand induced by HDV thus always outweighs their potential flexibility benefits. Cost effects, however, vary considerably between different options. Flexible BEV with V2G incur the lowest additional power sector costs of 1.8~bn~Euros/year (around 5,600~Euros/year per vehicle), followed by optimally charged BEV without V2G with 2.3~bn~Euros/year (around 7,200~Euros/year per vehicle). If BEV charging is not optimized, system costs are markedly higher, with 3.8~bn~Euros/year (11,900~Euros/year per vehicle). Results are qualitatively similar for ERS-BEV, but on a slightly higher cost level. The differences between the three ERS-BEV cases are less pronounced than for pure BEV, as their temporal flexibility potential is much smaller. The battery capacity of an ERS-BEV is only around a quarter of that of a pure BEV (655~kWh usable capacity per truck vs. 181~kWh), so ERS-BEV can make less use of hours with low electricity prices (compare also Figure~\ref{fig_si11}). In contrast, power sector costs are higher for FCEV (12.6 or 12.7~bn~Euros/year, for decentralized or centralized hydrogen provision, i.e.~around 39,700~Euros/year per vehicle) and even more so for PtL (16.8~bn~Euros/year, or 52,700~Euros/year per vehicle). This is a consequence of high conversion losses of hydrogen and PtL supply chains and vehicles drive trains. Because of these losses, the two hydrogen supply chains increase the electricity demand more than twice as much as the battery-electric options. The electricity demand of PtL-HDV is nearly four times as high as in the case of BEV. Notably, the cost differences between BEV and ERS-BEV scenarios are much smaller than the differences between these direct-electric options and indirect electrification via hydrogen or PtL.

We also evaluate average yearly wholesale electricity prices for HDV electricity (Figure~\ref{fig_02}, lower panel).\footnote{We interpret the dual variables of the model's energy balance as hourly wholesale prices, which is common practice in energy modeling \cite{brown_decreasing_2021,lopezprol_economics_2021}.} This allows to separate the differences in overall electricity consumption of the various HDV options from their ability to make use of low-cost electricity. Average prices are calculated by multiplying hourly wholesale prices of electricity consumed by the different HDV options or fed back to the grid with respective hourly quantities, summing up over the whole year, and dividing by the overall electricity consumption of the fleet. That is, the numbers also account for revenues of electricity sold via V2G, assuming that HDV operators receive the respective hourly wholesale price whenever they feed electricity back to the grid. BEV with V2G accordingly face the lowest average electricity prices, as these also benefit from revenues of feeding back to the grid, followed by BEV without V2G.

Average electricity prices paid by ERS-BEV are slightly higher, as their smaller batteries limit the ability to optimize their charging and V2G decisions. In contrast, pure BEV can leverage their larger batteries to make better use of hours with low electricity prices. In the case of inflexible charging, average electricity prices faced by inflexibly charged BEV are high, and even slightly above those of inflexible ERS-BEV. This is because inflexibly charged BEV benefit less from cheap electricity prices around midday related to abundant PV feed-in than inflexible ERS-BEV, which on average have a better (catenary) grid connection during these hours.

Prices for electricity used in hydrogen and PtL supply chains are in the same range as those of ERS-BEV options but below those of inflexibly charged BEV or ERS-BEV. Centralized hydrogen and PtL supply can make use of lower prices than decentralized supply, because their low-cost storage options allow for higher temporal flexibility. Electricity prices of centralized hydrogen supply chains and PtL are also cheaper than those faced by inflexibly operated BEV or ERS-BEV. In terms of overall costs, these temporal flexibility benefits are, however, by far outweighed by the higher overall energy consumption of the FCEV and PtL options (Figure~\ref{fig_02}, upper panel).

The average wholesale price in the reference scenario without electrified HDV is 84.9~Euros/MWh. Inflexibly charged BEV and ERS-BEV, as well as inflexible on-site electrolyzers, face slightly above-average prices. In contrast, the more flexible HDV options can make use of below-average market prices.

\subsection{Capacity and dispatch effects}\label{sec: capacity and dispatch effects}

\begin{figure}[ht!]
    \begin{mdframed}
        \centering
        \includegraphics[width =0.9\textwidth]{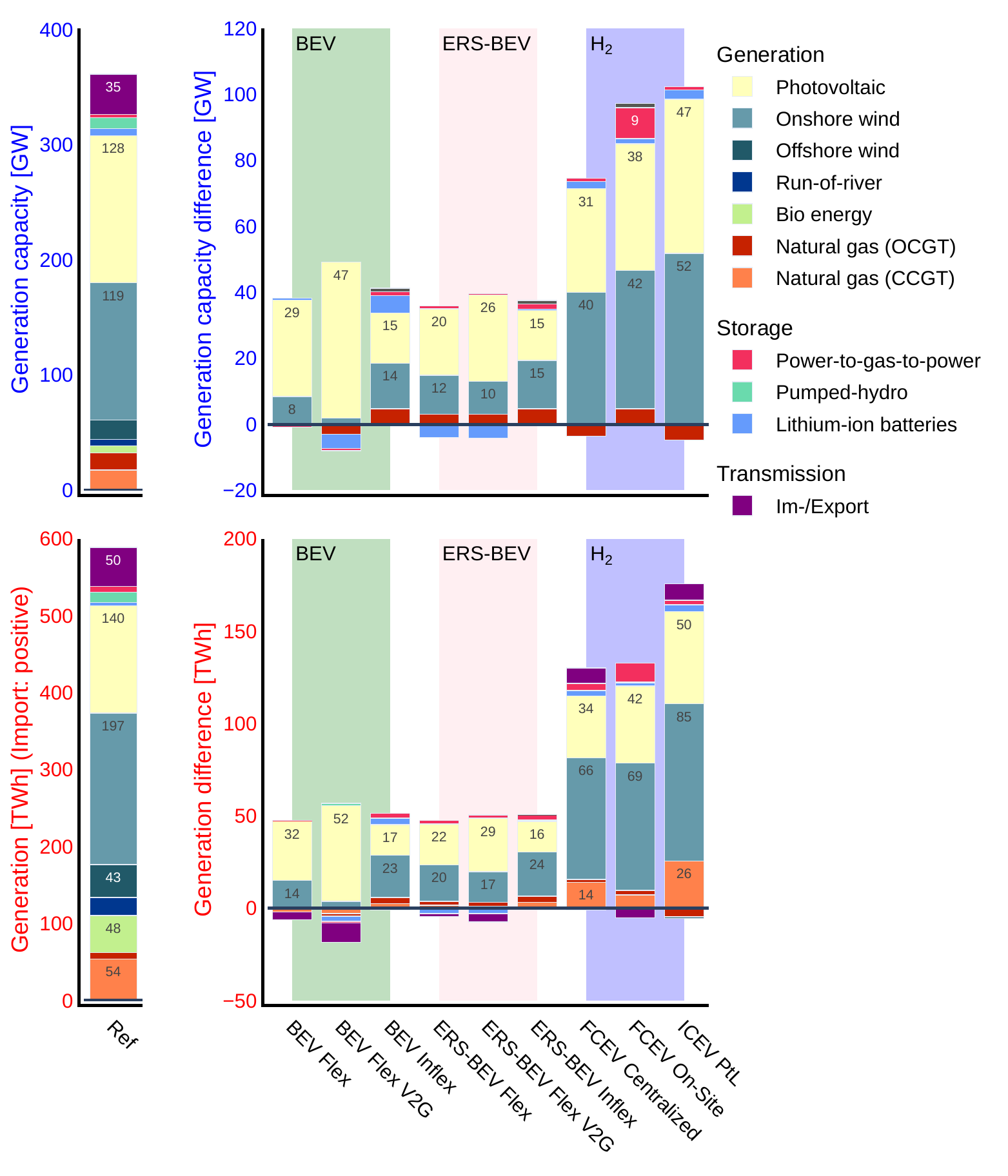}
    \end{mdframed}
    \caption{Effects of different HDV scenarios on optimal generation capacity (upper panel) and on yearly generation (lower panel) in Germany.}
    \label{fig_03}
\end{figure}

The upper panel of Figure~\ref{fig_03} shows optimal generation capacities in the reference (on the left) and the changes induced by HDV (on the right) in Germany. In the reference, variable solar PV and onshore wind power dominate the capacity mix.\footnote{In the Supplemental Information, we discuss the effects of including an upper bound for wind power investments by the year 2030, see section~\ref{sec: sensitivity wind}.} These are complemented by smaller firm capacities of natural gas and bioenergy. Optimal capacity additions related to the electrification of HDV are predominantly a mix of solar PV and onshore wind power, as these are the lowest-cost options available. Flexible BEV and ERS-BEV lead to the highest PV shares in the capacity additions, especially if combined with V2G. BEV trucks with V2G serve as a short-duration electricity storage option with a low energy-to-power ratio and many yearly cycles, which favors the expansion of solar PV. The potential for solar PV integration via V2G is less pronounced for ERS-BEV than for BEV, because ERS-BEV have smaller batteries.

HDV options that are temporally less flexible or that require more electricity favor higher onshore wind power capacities. In this case, the wind power generation profile fits better to HDV electricity demand than solar PV. Offshore wind power is not added here because of unfavorable cost assumptions compared to onshore wind power.\footnote{Given new policy targets and the outcomes of current tenders for offshore wind power capacity in Germany, offshore wind power may play a larger role than assumed here.} FCEV and PtL options have the highest capacity needs because of substantial conversion losses. On-site electrolysis further requires an addition of long-duration electricity storage capacity (power-to-gas-to-power) of 9.4~GW and 6.0~TWh (see section~\ref{sec: si storage}, Figure~\ref{fig_si06}). This is necessary to compensate for the temporal inflexibility of on-site electrolyzers (Figure\ref{fig_si07}) without large-scale hydrogen storage, which are forced to operate with a similar time profile as hydrogen demand. Among inflexibly charged battery-electric HDV options, BEV require larger investments in stationary lithium-ion batteries than ERS-BEV. Largely due to a CO$_{2}$ price of 100~Euros/ton, hard coal, lignite or oil generators are not used in any of the scenarios, while natural gas (CCGT) capacity is always at the assumed upper limit of 17.6~GW (compare section~\ref{sec: exp proc}). Up to 5~GW of open cycle gas turbines (OCGT) are added in some scenarios, especially for HDV options with low temporal flexibility (BEV Inflex, ERS-BEV Inflex and FCEV On-Site). 

Yearly electricity generation in the reference, as well as the changes induced by different HDV scenarios, are shown in the lower panel of Figure~\ref{fig_03}. 
The picture is generally similar to optimal capacity changes, but the share of wind power in additional electricity generation tends to be higher because of higher full-load hours as compared to solar PV. Flexible BEV, particularly if combined with V2G, cause the highest increase in PV generation, which can increase the renewable share of the German power sector to nearly 84\% (section~\ref{sec: renewable shares}, Figure~\ref{fig_si08}). Centralized hydrogen or PtL supply chains, which require much more electricity than BEV and ERS-BEV, use their temporal flexibility to also increase electricity imports and the use of the most energy-efficient natural gas-fired power plants (CCGT). Note that this flexibility comes from the option of low-cost hydrogen or PtL storage. This is different in the case of temporally inflexible hydrogen production at filling stations (FCEV On-Site). As imports and CCGT are less available in the hours of decentralized hydrogen production, more wind and solar generation in combination with long-duration electricity storage are used, which also increases the share of renewables in the overall system to nearly 83\%. Higher utilization of natural gas also translates into increasing direct carbon emissions (section~\ref{sec: CO2}, Figure~\ref{fig_si09}). The effects of BEV and ERS-BEV on CO\textsubscript{2} emissions are generally lower, and may even become negative when combined with V2G. This is because V2G allows for an increase in solar photovoltaic capacity and a higher renewable energy share, which can slightly reduce overall carbon emissions compared to the reference case without electrified HDV.

\subsection{Time series reveal differences in flexibility characteristics}

\begin{figure}[htp!]
    \begin{mdframed}
        \centering
        \includegraphics[width =0.96\textwidth]{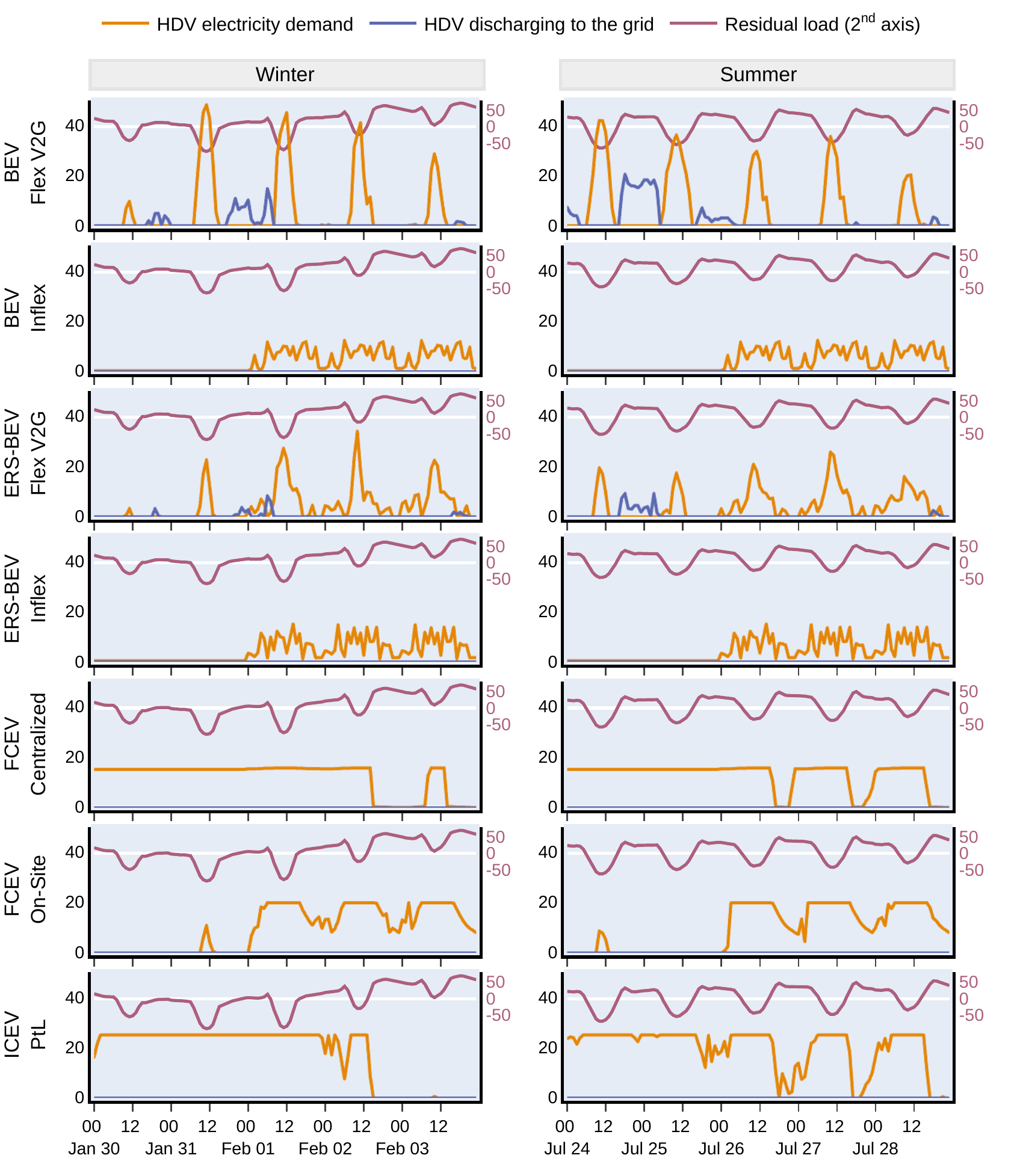}
    \end{mdframed}
    \caption{Electricity time series sample of five exemplary days (unit: GW): ``HDV electricity demand'' illustrates electricity flowing to BEV or ERS-BEV batteries, including amounts later charged back via V2G, or electricity needs of hydrogen and PtL supply chains; ``HDV discharging to the grid" are V2G power flows from BEV or ERS-BEV to the grid; and ``Residual load'' is the net load of the electricity sector after considering the feed-in potential of variable renewables. The samples start on a Saturday. For simplification, we assume that both Saturday and Sunday are completely truck-free. BEV Flex and ERS-BEV Flex without V2G are omitted in the graph, as charging patterns are very similar to the shown V2G cases. Additional time series are shown in the SI, Figures~\ref{fig_si10},~\ref{fig_si12},~\ref{fig_si13}.}
    \label{fig_04}
\end{figure}

Flexibly operated BEV trucks (BEV Flex V2G) are able to charge their batteries in hours of low residual load, as illustrated in Figure~\ref{fig_04}. Such periods generally go along with low wholesale market prices (compare Figure~\ref{fig_si11}). This enables BEV to utilize renewable surplus energy (i.e.,~negative residual load) to a substantial extent. They also make use of the V2G option and feed back renewable surplus energy to the grid, whenever the battery capacity is not needed for driving. This is particularly visible during summer during weekends, when trucks are not driving, but assumed to have a grid connection (top right panel in Figure~\ref{fig_04}). Here, BEV with V2G charge their batteries with solar PV surplus energy on Saturday afternoon, and feed it back to the grid in the night between Saturday and Sunday. This is different in the following days, as HDV are used between Monday and Friday, and much less battery capacity is available for V2G. If BEV charging is not optimized, but follows a pre-determined pattern (BEV Inflex), charging profiles are less peaky and much more balanced. BEV are not able to make particular use of cheap renewable surplus electricity in this case, but carry out a substantial part of their charging in hours with positive residual load. 

ERS-BEV follow similar patterns as non-catenary BEV. Yet, their smaller batteries make them temporally less flexible, so they can make less use of renewable surplus energy and have to draw more electricity from the grid during hours with positive residual load. Their potential for feeding electricity back to the grid is thus also much smaller than in the case of pure BEV. ERS-BEV sometimes also charge their batteries during driving (compare Figure~\ref{fig_si12}). 

Hydrogen and PtL supply chains show very different patterns of electricity use. Centralized hydrogen supply chains (FCEV Centralized), which come with large-scale hydrogen storage capacities, generally have a flat consumption profile in many hours, as their high fixed costs make it optimal to use them with high full load hours. This limits their ability to make use of renewable surplus energy. Yet, they can use the temporal flexibility provided by large-scale hydrogen storage to reduce electricity consumption in hours of high positive residual load, i.e.,~high prices. Decentralized hydrogen supply chains (FCEV On-Site) have an electricity use profile that largely follows the hydrogen demand profile, driven by limited on-site hydrogen storage capacity. Accordingly, their potential for avoiding electricity consumption in hours of high residual load is lower. The PtL supply chain (ICEV PtL) also comes with large-scale storage and thus has a similar pattern as the centralized hydrogen option. The load, however, is generally higher because of higher overall energy consumption of PtL versus hydrogen. This goes along with larger investments in electrolyzers, which are 24.4~GW in the ICEV PtL scenario, as compared to 13.5~GW for FCEV Centralized (section~\ref{sec: electrolysis capacity}).

\section{Discussion}\label{sec: discussion}

\subsection{Limitations and avenues for future research}
Our model-based analysis of the power sector effects of alternative options for de-fossilizing heavy-duty vehicles inevitably has a range of limitations. We briefly discuss how these may qualitatively impact our results.

First, we deliberately focus on a detailed analysis of the power sector implications of alternative HDV electrification options to address this particular research gap. Unlike other studies, we do neither aim to compare total costs of ownership (\cite{nykvist_feasibility_2021,hunter_spatial_2021,samet_2024}), nor the overall system costs of different options. Including the latter in our power sector-focused analysis would also require considering infrastructure costs related to catenary lines, chargers, or filling stations, as well as differences in vehicle costs. Such future costs are highly uncertain, so Monte Carlo approaches or other methods to deal with parameter uncertainty would have to be used. Note that the differences in the costs of vehicles and related charging infrastructure might in fact partially outweigh the differences in power sector costs identified here, especially the small differences between pure BEV and ERS-BEV. Yet, the very large power sector cost differences between (ERS-)BEV on the one hand, and FCEV or PtL trucks on the other, would require the latter to be substantially cheaper in terms of vehicle and infrastructure costs in order to break even, which appears unlikely.

While focusing on the power sector, we further assume extensive charging and overhead line infrastructure to be available for BEV and ERS-BEV, both in depots and en route, and that vehicles are always connected to the grid when idling. In reality, the availability of grid connections is likely to be lower for a variety of reasons, which would decrease the flexibility potentials of optimized charging. In sensitivity analyses, we show that decreasing charging availability at depots reduces the potential for flexible charging and particularly for vehicle-to-grid operations of BEV HDV (\ref{sec: sensitivity charging availability}). This appears to be in line with the finding that the flexibility benefits of very high levels of grid connection may not necessarily outweigh related infrastructure costs \cite{brinkel_2020}. In further sensitivities, we conversely find that higher assumptions on fast-charging power ratings do not have a major effect on results (Figures~\ref{fig_si20},~\ref{fig_si21},~\ref{fig_si22}).

Another simplification of our numerical analysis is the assumption of perfect foresight, which results in an upper limit of the flexibility benefits that can be realized by optimized (dis-)charging of BEV and ERS-BEV. We also abstract from including electric passenger vehicles or other future sector coupling options, which may provide flexibility to the power sector and compete with electric trucks for electricity use in times of high renewable availability. In particular, power-to-heat technologies may be able to efficiently exploit hours of high renewable generation and low energy prices, as heat storage is relatively cheap compared to electricity storage \cite{bloess_2018}. In contrast, we may underestimate temporal flexibility constraints of the PtL production process.

To facilitate numerically efficient solutions and keep results tractable, we further assume electricity generation capacities to be fixed in Germany's neighboring countries. At the same time, the modeled HDV fleets only cover domestic truck transport in Germany. This may lead to an overestimation of the power sector flexibility supplied by Germany's neighbors for integrating electric trucks in Germany. Likewise, increasing the spatial resolution within a country could raise additional insights. Taking into account transmission grid bottlenecks and differences in regional potentials for wind and solar power, it may be more beneficial to charge HDV trucks in some regions than in others. We further abstract from comparing different options for the distribution of hydrogen to filling stations, which has been done in detail in previous work \cite{stoeckl_2021}.

Future research may address these limitations. While our analysis focuses on the power sector, it also appears desirable to investigate complementary cost measures. This includes detailed total cost of ownership analyses from a HDV operator perspective, as well as investigations of overall system cost effects, which would also consider cost differences of charging and ERS infrastructures, as well as purchase cost differences of different types of HDV. This requires a whole set of additional and detailed parameter assumptions, which we leave for future research. Further, it would be desirable to carry out similar analyses for other countries and world regions. While we focus on Germany, we expect that general results should also hold for comparable non-hydro electricity sectors in other countries in temperate climates.

\subsection{Differences in temporal flexibility and energy efficiency drive effects}
Our analysis focuses on the effects of alternative options for de-fossilizing heavy-duty vehicles on power sector costs as well as optimal investment and dispatch decisions of electricity generation and storage technologies. We find that differences in temporal flexibility and energy efficiency are important drivers of results.

Fleets of BEV or ERS-BEV trucks that are charged in an optimized way cause the lowest increase in power sector costs. This is because they require much less electricity than HDV powered by electrolysis-based hydrogen or e-fuels. At the same time, they can offer temporal flexibility to the power sector, which lowers the costs compared to non-optimized charging. This is increasingly valuable with growing shares of variable renewable energy sources. Flexibility benefits are particularly large if BEV also carry out vehicle-to-grid operations, i.e.,~if the truck batteries can be used as a distributed grid storage option. In contrast, indirect electrification of the HDV fleet via FCEV or PtL incurs much higher power sector costs, even if compared to inflexibly charged BEV or ERS-BEV. That is, the temporal flexibility benefits of centralized hydrogen or PtL supply chains, which come with low-cost storage options, do not outweigh their energy efficiency drawbacks in terms of power sector costs.

Alternative options of (in-)directly electrifying HDV fleets further entail different optimal generation capacity mixes. The additional electricity for battery-electric HDV is largely supplied by solar PV in the lowest-cost solution, particularly if V2G is also available. In contrast, centralized hydrogen and PtL supply chains go along with higher optimal shares of wind power, and also lead to a higher use of non-renewable electricity generators. The latter translates into higher direct carbon emissions of these technologies. However, emission outcomes depend on the assumed boundary conditions, such as capacity bounds for non-renewable generators and CO$_{2}$ prices.

Additional sensitivity analyses show that results are generally robust. For instance, Germany is modeled as an electric island in a sensitivity analysis, i.e.,~the flexibility benefits of the European interconnection are neglected (section~\ref{sec: sensitivity island}, Figures~\ref{fig_si14},~\ref{fig_si15},~\ref{fig_si16}). In such a case, flexible BEV with V2G are even more beneficial, and inflexible BEV operations lead to slightly higher cost increases because an isolated power sector is more flexibility-constrained. Further, results are robust against the assumption that the potential for wind power expansion is limited, and solar PV accordingly plays a larger role (section~\ref{sec: sensitivity wind}, Figures~\ref{fig_si17},~\ref{fig_si18},~\ref{fig_si19}).

\subsection{Direct electrification and smart charging preferable}
From a pure power sector perspective, direct electrification of the truck fleet would thus be clearly preferable. Aside from the cost benefits determined here, this appears to be even more important as the utilization of renewable energy sources is unlikely to grow exponentially \cite{hansen_2017}, and renewable growth rates in fact have been smaller than what would be required in $1.5^{\circ}C$-compatible scenarios \cite{cherp_2021}. A similar argument has been made for the global scale-up of electrolysis \cite{odenweller_2022}. All of this calls for energy-efficient, direct electrification options of HDV wherever possible.

Moreover, smart charging of BEV or ERS-BEV is highly desirable, as it provides temporal flexibility which leads to lower power sector costs and CO\textsubscript{2} emissions, compared to non-optimized charging. This corroborates the findings of Pickering et al., who also highlight the benefits of controlled vehicle charging in decarbonized and largely renewable future European energy systems \cite{pickering_diversity_2022}. If feeding back electricity to the grid via V2G is possible, this further lowers the costs and increases the optimal shares of solar PV.

\clearpage

\section{Experimental procedures}\label{sec: exp proc}

\subsection{Resource availability}

\subsubsection{Lead contact}

Further information and requests for resources should be directed to the lead contact, Wolf-Peter Schill (wschill@diw.de)

\subsubsection{Materials availability}

This study did not generate new unique reagents.

\subsubsection{Data and code availability}

All data and code are available under permissive licenses in public repositories. The general repository of the DIETER model can be found here: \url{https://gitlab.com/diw-evu/dieter_public}. The particular model version used for the present study, together with the complete input data set (including BEV and ERS-BEV profiles), is available under permissive licenses in a dedicated repository. This also includes model results in Jupyter notebook and html formats: \url{https://gitlab.com/diw-evu/projects/my_eroads}. A general model documentation is available here: \url{https://diw-evu.gitlab.io/dieter_public/dieterpy/}.

\subsection{The power sector model DIETER}
We use the open-source power sector model \textit{Dispatch and Investment Evaluation Tool with Endogenous Renewables} (DIETER). It is a linear program that minimizes power sector costs by optimizing capacity and dispatch decisions for a full year in an hourly resolution \cite{zerrahn_long-run_2017,gaete_2021}. Its objective function includes fixed an variable costs of all electricity generation and storage technologies, electrolysis and PtL plants, as well as hydrogen or e-fuels transportation. It does not include the costs of charging or catenary infrastructure, hydrogen filling stations, or PtL filling stations. Accordingly, the power sector cost figures provided above do not include the costs of HDV electrification infrastructure. We further do not consider the option of hydrogen imports, as these are likely to be unavailable at scale by 2030. In general, the global scaling up of green hydrogen supply remains uncertain \cite{odenweller_2022}. 

Endogenous model variables include power sector costs, optimal generation and electricity storage capacities (Germany) and their hourly use (all countries), hourly decisions for HDV charging and discharging, as well as the capacity and operational decisions of electrolysis and PtL generation and storage infrastructure. In addition, we interpret the marginals of the hourly energy balance as wholesale prices \cite{brown_decreasing_2021}. 

Exogenous model inputs include fixed and variable costs of all electricity generation and storage technologies, efficiency parameters, as well as time-series of variable renewable energy availability and electric load. We further include HDV driving profiles as explained in the following section~\ref{sec: traffic data} and section~\ref{sec: BEV profiles}. In the case of inflexible HDV charging (BEV Inflex, ERS-BEV Inflex), we assume that the vehicles always start charging as soon as an opportunity arises, and that vehicles batteries are fully charged by the time the next trip starts, if possible. That is, the charging power is lower, the longer a vehicle is connected to the grid. This resembles the ``balanced'' charging profile defined in \cite{gaete_2021_emobpy}.

The geographic scope of the model version used here includes Germany and its neighboring countries plus Italy. In order to reduce numerical complexity and improve tractability, we allow for endogenous generation capacity investment only in Germany, and fix the power plant portfolio for the other countries to values derived from ENTSO-E's Ten Year Network Development Plans \cite{tyndp_2018,tyndp_2020}. We further assume upper limits for investments into fossil generation technologies according to the main scenario for 2030 from the German Federal Grid Development plan (NEP, \cite{nep_2018}). 

Different versions of the model have been used in various earlier studies to investigate various aspects of renewable energy integration, electricity storage, and sector coupling \cite[e.g.,~][]{schill_2018,schill_joule_2020,stoeckl_2021,gils_2022}. The DIETER model is available open-source.

\subsection{Traffic data of heavy-duty vehicles}\label{sec: traffic data}

We generate synthetic truck usage patterns that are intended to approximate the German fleet of HDV larger than 26~tons. The main data source for the usage patterns is the traffic model PTV Validate \cite{ptv_2023}, from which we extract a database of daily truck trips in domestic German road freight transport. In order to derive typical daily driving profiles, the following steps are performed (for more detailed information, see~SI):

\begin{enumerate}
    \item Calculation of annual mileage for transport relations. For this step, we build on statistics of typical annual mileages for different distance classes and derive a steady function. 
    \item Calculation of daily idle time. This is calculated based on the number of daily truck trips (and the number of stops in between) as well as typical times for loading and unloading, depending on the goods type. In addition, a mandatory driver’s break of 45~minutes is assumed for daily driving profiles \textgreater{} 4.5 hours.
    \item The daily operating time of a vehicle on a particular route is determined from the above assumptions using the formula ``operating time = journey time + idle time + driver’s break''.
\end{enumerate}

Repeating steps 1-3, synthetic daily travel profiles with a certain daily operating time are obtained for all transport relations in the model. In order to limit these to a manageable number of daily driving profiles for the power sector model, all relations that have approximately the same daily operating time are combined. We derive a number of 19 daily profiles which are distributed over the course of the day such that the empirically observed course of the daily mileage on the German road network is reproduced. For these profiles, the time series of charging availability (in the depot, during idle and driver’s resting times) and of the electricity demand of pure BEV-HDV (500~km battery range) and ERS-HDV (150~km battery range) are calculated. The resulting electricity demands and charging availabilities are used as inputs for the DIETER model.

For FCEV and PtL trucks, only aggregate fuel demand time series are required. These are based on the same empirical traffic data from German highways \cite{bast_2023}. For FCEV, we assume a specific hydrogen demand of 6.8~kg/100km, and for ICEV-PtL a specific Diesel demand of 27.1~l/100km.

\section*{Acknowledgments} 
We thank Hinrich Helms and Michel Allekotte for their valuable comments on earlier drafts. This work has benefited from research grants by the Federal Ministry for the Environment, Nature Conservation and Nuclear Safety (BMU) and by the Federal Ministry for Economic Affairs and Climate Action via the projects ``My eRoads'' (Fkz 16EM4006-1) and ``enERSyn'' (Fkz 01MV22004B).

\section*{Author contributions}
Conceptualization: JJ, WPS; Methodology: CG, FH, JJ, WPS; Software: CG; Investigation: CG, FH, JJ, WPS; Data curation: FH, WPS; Writing - original draft: CG, WPS; writing - review \& editing FH, JJ; Visualization: CG, FH; Project administration: JJ; Funding acquisition: JJ, WPS.

\section*{Declaration of interests}
The authors declare no competing interests.


\defaultbibliographystyle{elsarticle-num}
\biboptions{super,sort&compress}
\putbib
\end{bibunit}

\clearpage

\begin{bibunit}

\renewcommand{\thesection}{S}
\renewcommand{\thepage}{S}
\global\long\def\thefigure{S.\arabic{figure}}
\global\long\def\thetable{S.\arabic{table}}
\global\long\def\thepage{S.\arabic{page}}
\setcounter{figure}{0}
\setcounter{table}{0}
\setcounter{page}{1}
\newpage

\section{Supplemental Information}

\subsection{Supplemental Experimental Procedures}

\subsubsection{Generation of BEV and ERS-BEV profiles}\label{sec: BEV profiles}
Vehicle load profiles are key input parameters to this analysis. In case of non-optimized, inflexible charging, respective time series are provided to the power sector model as exogenous parameters. In case of optimized charging, the vehicles' charging and, if V2G is available, discharging profiles are determined as endogenous variables. In this case, mission-profiles become important model inputs. Data sets for truck trips are available for Germany and also Europe \cite{speth_synthetic_2022}. However, these trips are not connected to vehicle mission profiles, e.g. information on daily starting and charging times, in these data sets. Therefore, previous studies often refer to the observed traffic volume on typical roads in order to approximate the energy demand profile of the vehicle fleet \cite{taljegard_impacts_2019}. However, it is impossible to determine flexibility potentials in the load profile when using this approach.

This study is based on synthetic truck usage patterns that are intended to roughly approximate the German HDV fleet (\textgreater{}~26 t) in 2030, when a broad electrification of heavy-duty road transport is assumed. The main data source for the usage patterns is the traffic model PTV Validate \cite{ptv_2023} which in turn is based on the official German governmental forecast for goods flow matrices for the year 2030 \cite{schubert_2014}. Within PTV Validate, the freight flows are allocated to vehicle trips between approximately 10,000 origin and destination districts in Germany, i.e.,~to individual transport relations. For the present study, we use aggregated source and destination districts at county level (approx.~400 counties in Germany). PTV Validate also determines the vehicle class of trucks used for the transport of a certain goods type on a given transport relation. Only the journeys of trucks in the size class~\textgreater{}~26 t are considered here, as these account for a large proportion of the energy consumption of the truck fleet and also represent the primarily intended field of application for electricity consumption by overhead lines.

The result is a database of daily trips by HDV in domestic German traffic. International traffic is not considered here to reduce complexity. Note that the purpose of this analysis is mainly to illustrate the general power sector implications of different HDV technologies, rather than forecasting their absolute impacts. 

In order to derive typical daily driving profiles from the aforementioned trip database, the following steps are performed:

\begin{enumerate}
    \item \textbf{Calculation of annual mileage for transport relations:} The distance of a transport relation is assumed as characteristic for the radius of operation that a vehicle performing transports on the given transport relation will typically have. This allows us to utilize official statistics from the German Federal Road Administration which give typical annual mileages per vehicle class for the operating ranges 0-50 km, 50-150 km and \textgreater{}~150 km. Using a regression approach, a dependency between the distance of each transport relation and the typical annual mileage of vehicles operating there is obtained.
    \item \textbf{Calculation of daily idle time:} First, the number of daily trips made by a truck is estimated. For example, trucks on short shuttle routes often make multiple trips a day, whereas trucks on long-distance routes often only make one trip a day. The number of trips entails a corresponding number of stops, which can in principle be used to charge the traction battery for the duration of the stop. Second, depending on the type of goods transported, typical durations for loading and unloading were estimated. These determine the duration of the assumed idle times in the daily driving profile. In addition, a mandatory driver’s break of 45 minutes is assumed for daily driving profiles of \textgreater{}~4.5 hours.
    \item \textbf{Calculation of daily operating time:} The daily operation time of a vehicle on a particular route is therefore defined as 
    \begin{equation}
        \textrm{operating time = journey time + idle time + driver’s break.}
    \end{equation}

\end{enumerate}

For step 1, we use the regression function

\begin{equation} 
M= a \cdot R^b + c
\end{equation}

where M is the annual mileage, R is the distance of a relation and a, b and c optimization coefficients. The objective function is constructed such that the average annual mileages of each distance class from Table~\ref{Table_a1} are reproduced when iterating over all transport relations.

\begin{figure}[ht!]
    \centering
    \includegraphics[width=12cm]{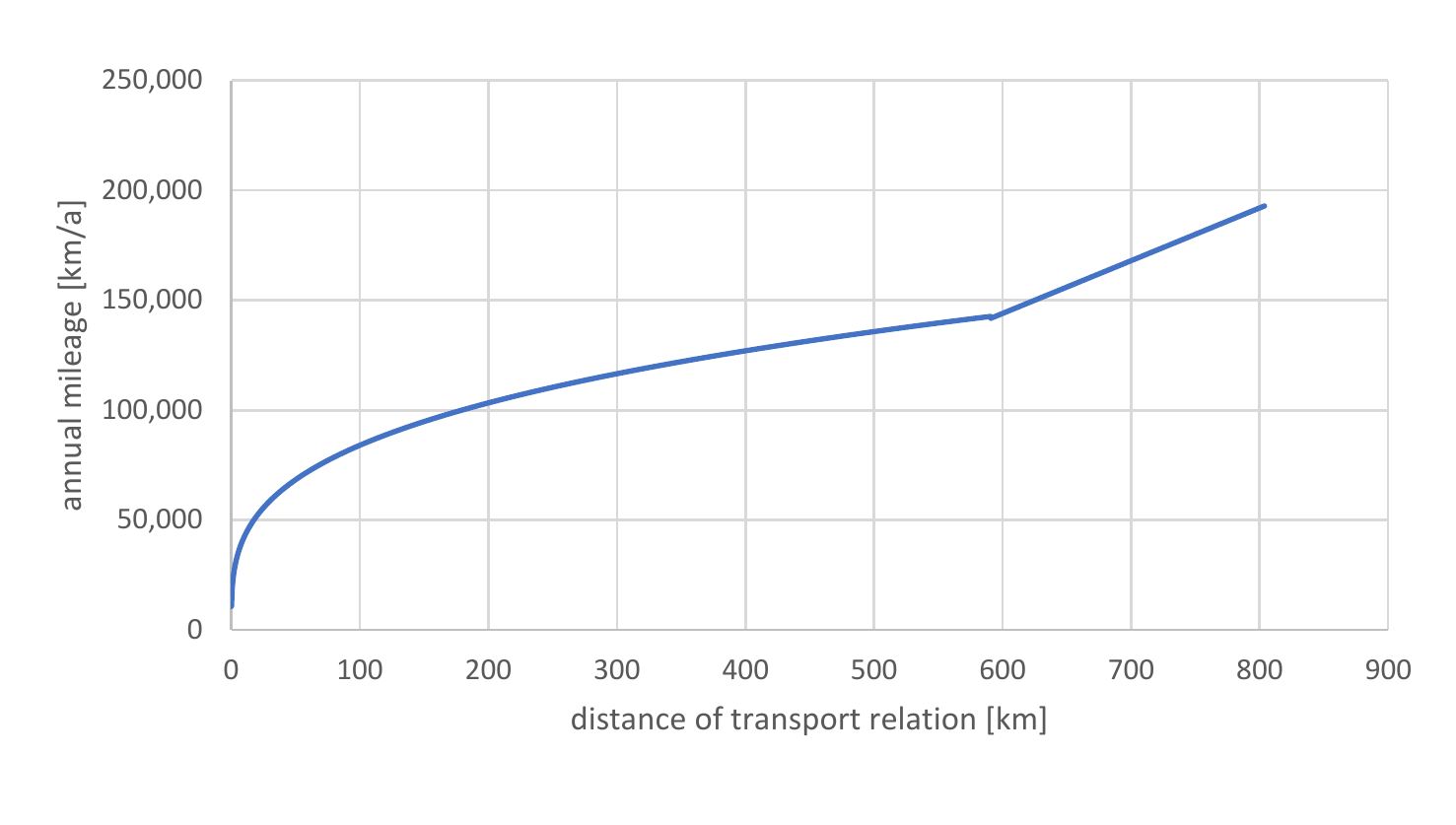}
    \caption{Annual mileage as a function of distance class.}
    \label{fig_si01}
\end{figure}

The regression result for the parameters is a~=~21,180, b~=~0.2979 and c~=~0. The resulting dependency is shown in Figure~\ref{fig_si01}.

\begin{table}[ht!]
    \caption{Annual mileage of HDV \textgreater{}~26~t per distance class. Source: \cite{bast_2017}}
    \label{Table_a1}
    \centering
    \begin{tabular}{|c|c|}
    \hline
    Distance class      & \begin{tabular}[c]{@{}c@{}}Average annual mileage \\ {[}km{]}\end{tabular} \\ \hline
    \textless 50 km     & 45,684                                       \\
    51-150 km           & 78,190                                       \\
    \textgreater 150 km & 117,121                                      \\
    All classes         & 94,950                                       \\ \hline
    \end{tabular}
\end{table}

For step 2, we use the ratio of the daily mileage (i.e.,~annual mileage (cf.~step 1), divided by 240 working days) and the distance of the transport relation to obtain the number of trips per day and thus the number of stops on a daily mission. The idle time per stop is determined based on an expert guess for each goods class. By multiplying the number of stops with the idle time per stop, we obtain the total daily idle time for the vehicle operation on a given transport relation, considering the type of transported goods.

Repeating steps 1-3, synthetic daily travel profiles with a certain daily operating time are obtained for all transport relations in the model. In order to limit these to a manageable number of daily driving profiles that can be included in the power sector model, all relations with a similar daily operating time are combined to eight stylized profiles (all relations with an operating time between 7.5 and 8.5~hours were aggregated to the 8-hour stylized profile). For each stylized profile, averaged values for the idle times are used. The essential parameters for the operational profiles are summarized in Table~\ref{Table_a2}. Note that the total annual mileage of each profile is the product of average daily mileage, 240 days of operation per year, and the number of vehicles.

\begin{table}[ht!]
\caption{Key data of daily driving profiles}
\label{Table_a2}
\centering
\resizebox{\columnwidth}{!}{%
\begin{tabular}{|c|c|c|c|c|c|}
\hline
\begin{tabular}[c]{@{}c@{}}Daily operating \\ time {[}h{]}\end{tabular} & \begin{tabular}[c]{@{}c@{}}Average distance of \\ relation {[}km{]}\end{tabular} & \begin{tabular}[c]{@{}c@{}}Average daily \\ mileage {[}km{]}\end{tabular} & \begin{tabular}[c]{@{}c@{}}Average daily \\ idle time {[}h{]}\end{tabular} & \begin{tabular}[c]{@{}c@{}}Number of \\ vehicles\end{tabular}  & \begin{tabular}[c]{@{}c@{}}Total annual \\ mileage {[}bn. km{]}
\end{tabular} \\ \hline
3  & 23  & 196  & 0.11 & 21,073 & 0.99   \\
4  & 74  & 281  & 0.06 & 29,007 & 1.96  \\
5  & 136 & 354  & 0.04 & 7,069  & 0.6   \\
6  & 175 & 383  & 0.78 & 34,772 & 3.2   \\
7  & 237 & 415  & 1.43 & 43,544 & 4.34   \\
8  & 219 & 399  & 2.43 & 86,826 & 8.31   \\
9  & 204 & 351  & 3.97 & 77,117 & 6.49   \\
10 & 448 & 527  & 2.89 & 19,166 & 2.42  \\ \hline
\end{tabular}%
}
\end{table}

The assumptions regarding the vehicles and their infrastructure are made independently of the daily operating time for the sake of simplicity and are summarized in Table~\ref{Table_a3}. The design of the vehicles is oriented towards the needs of long-haul transport with a range of 500~km for BEV. For profiles in local and regional transport (with mostly lower daily mileage), smaller batteries could generally also be sufficient, so the battery capacity tends to be overestimated in this analysis.

\begin{table}[ht!]
\caption{Assumptions on BEV and ERS-BEV HDV and infrastructure}
\label{Table_a3}
\centering
\resizebox{\columnwidth}{!}{%
\begin{tabular}{c|cc|}
\cline{2-3}
    & \multicolumn{1}{c|}{\textbf{BEV}} & \textbf{ERS-BEV} \\ \hline
\multicolumn{1}{|c|}{Vehicle size class} & \multicolumn{2}{c|}{\textgreater 26 t gross vehicle weight} \\ \hline
\multicolumn{1}{|c|}{Battery range}   & \multicolumn{1}{c|}{500 km} & 150 km        \\ \hline
\multicolumn{1}{|c|}{Effective battery capacity} & \multicolumn{1}{c|}{655 kWh} & 181 kWh \\ \hline
\multicolumn{1}{|c|}{Average speed}  & \multicolumn{2}{c|}{79 km/h} \\ \hline
\multicolumn{1}{|c|}{Energy consumption}  & \multicolumn{1}{c|}{1.31 kWh/km} & \multicolumn{1}{l|}{\begin{tabular}[c]{@{}l@{}}Overhead line connected:      1.42 kWh/km \\ Overhead line disconnected: 1.25 kWh/km\end{tabular}} \\ \hline
\multicolumn{1}{|c|}{\begin{tabular}[c]{@{}c@{}}Intermediate charging during \\ (un-)loading stop\end{tabular}}   & \multicolumn{1}{c|}{\begin{tabular}[c]{@{}c@{}}200 kW nominal\\ (166 kW effective )\end{tabular}} & \textbf{-} \\ \hline
\multicolumn{1}{|c|}{\begin{tabular}[c]{@{}c@{}}Intermediate charging during \\ mandatory driver's breaks\end{tabular}} & \multicolumn{1}{c|}{\begin{tabular}[c]{@{}c@{}}500 kW\\ (415 kW effective)\end{tabular}}   & \textbf{-}  \\ \hline
\multicolumn{1}{|c|}{Power supply from overhead line}  & \multicolumn{1}{c|}{-} & 400 kW \\ \hline
\end{tabular}%
}
\end{table}

\begin{figure}[ht!]
  \centering
  \subfloat{\includegraphics[width=\textwidth]{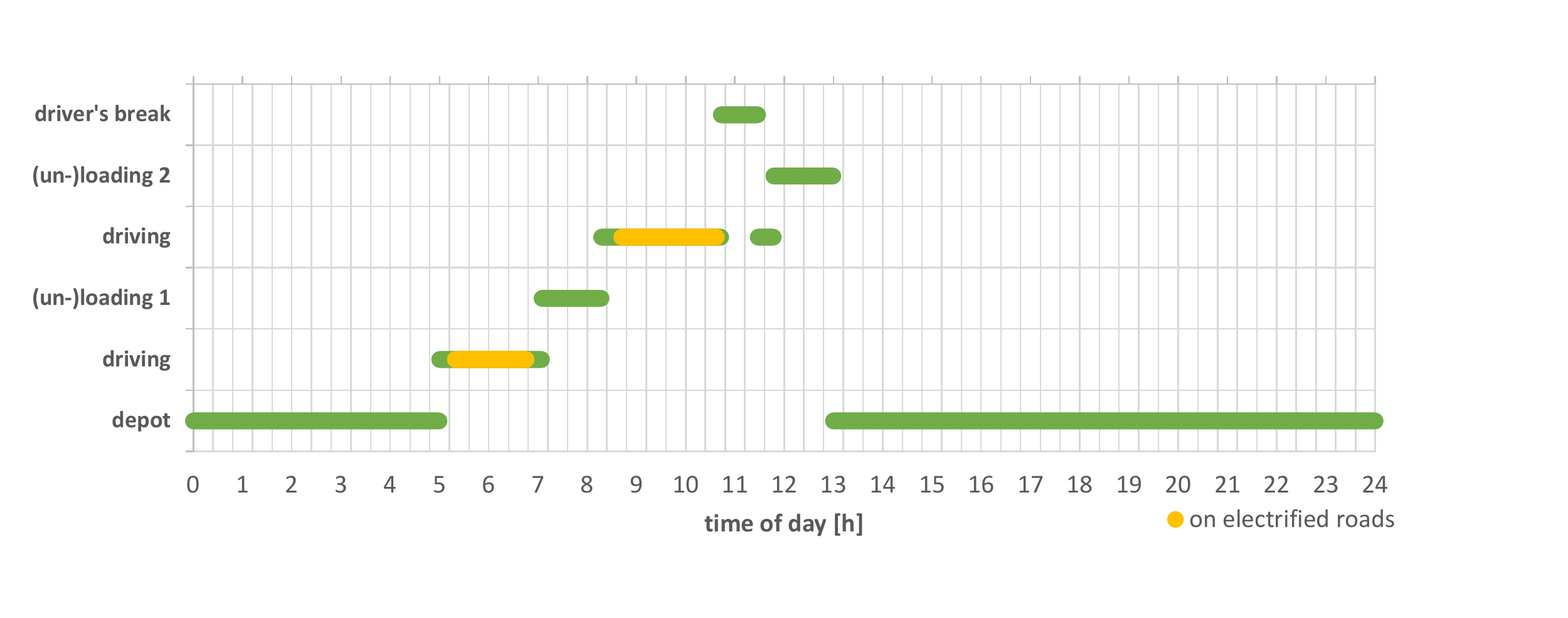}} \\ \centering
  \caption{Transport relations of an exemplary daily driving profile with an operating time of 8 hours}
  \label{fig_si02}
\end{figure}

The structure of the synthetic daily driving profiles is illustrated below using the example of a profile with an operating time of 8~hours (Figure~\ref{fig_si02}). The profile starts with a fully charged battery in the depot. For the case of ERS trucks, the share of electrified roads is assumed according to the motorway share of the transport relations on which the profile is based (assuming that all motorways are equipped with overhead lines). At loading/unloading stops, as well as in depots, a charging opportunity with a power rating of 200~kW is assumed for battery trucks, and a high-power charging option with 500~kW is assumed to be available during the mandatory driver’s break. For ERS trucks, on the other hand, a stationary charging opportunity is assumed only at the depot, with a similar power rating of 200~kW.

The number of idle stops during loading/unloading is in the range between 2 and 3 for average daily mileages above 300~km, which makes up the bulk of the total vehicle fleet. Only for short transport relations, it is higher. For the sake of simplicity, we assume 2 idling stops for all synthetic driving profiles. However, the calculation of absolute idle time per day (which is more relevant for the energy system modelling) is based on the respective number of idle stops for all contributing transport relations.

\begin{figure}[ht!]
  \centering
  \subfloat{\includegraphics[width=0.938\textwidth]{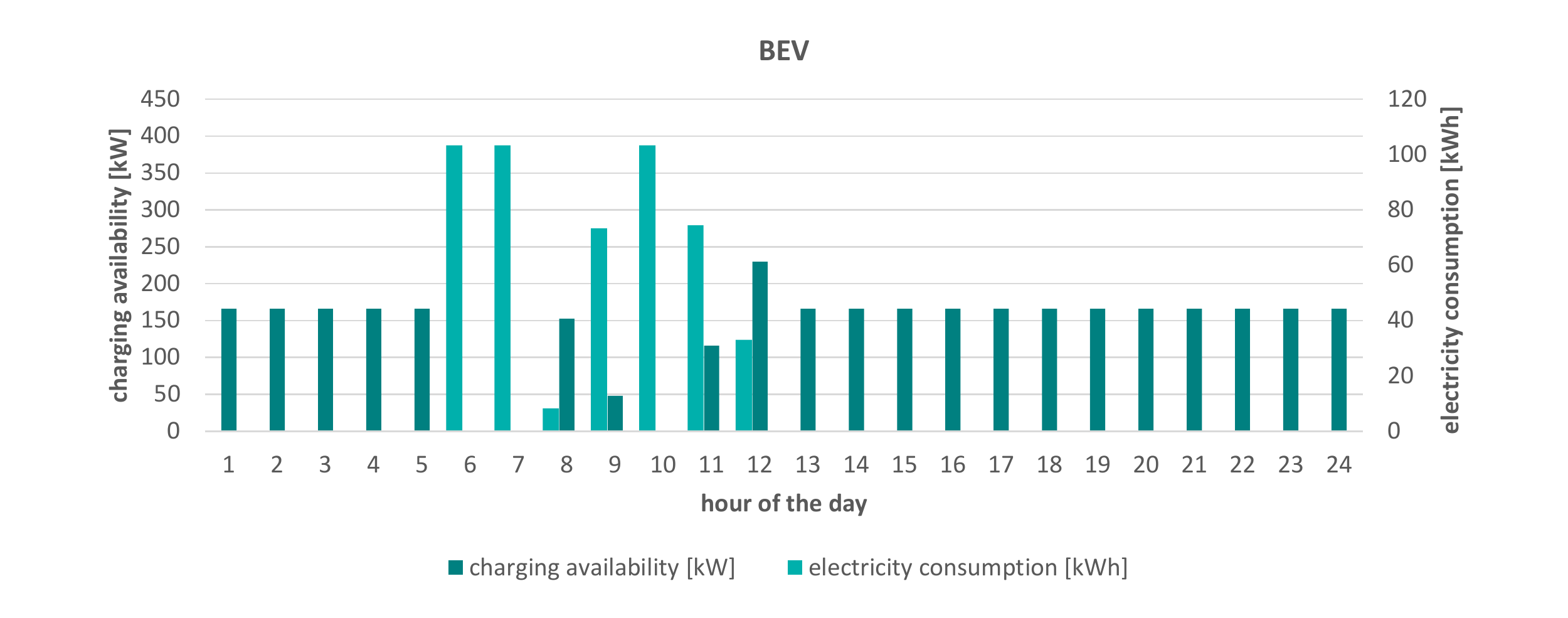}} 
  \caption{Resulting time-series for BEV of an exemplary daily driving profile with an operating time of 8 hours}
  \label{fig_si03}
\end{figure}

\begin{figure}[ht!]
  \centering
  \subfloat{\includegraphics[width=0.938\textwidth]{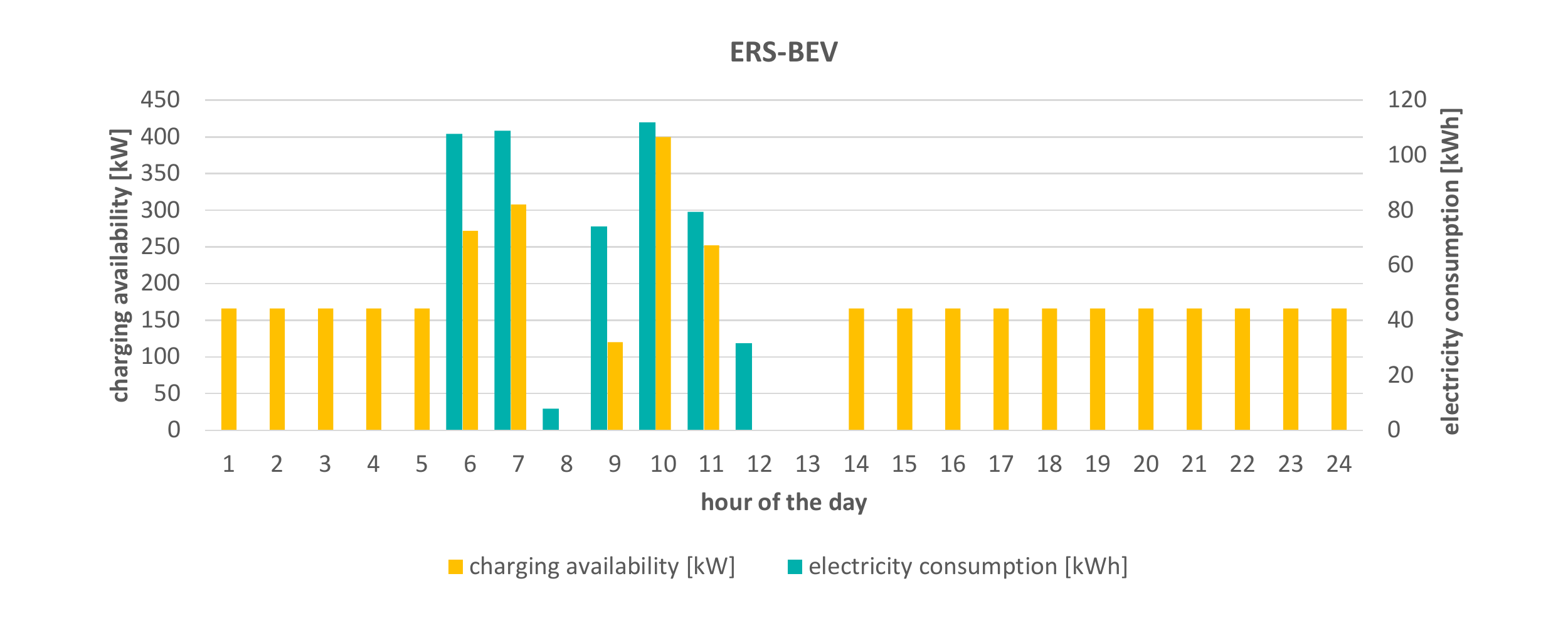}} 
  \caption{Resulting time-series for ERS-BEV of an exemplary daily driving profile with an operating time of 8 hours}
  \label{fig_si04}
\end{figure}

The resulting time series of hourly electricity demand and charging availability are shown in Figure~\ref{fig_si03} and Figure~\ref{fig_si04} for BEV and ERS-BEV respectively.

\begin{figure}[ht!]
    \centering
    \includegraphics[width=\textwidth]{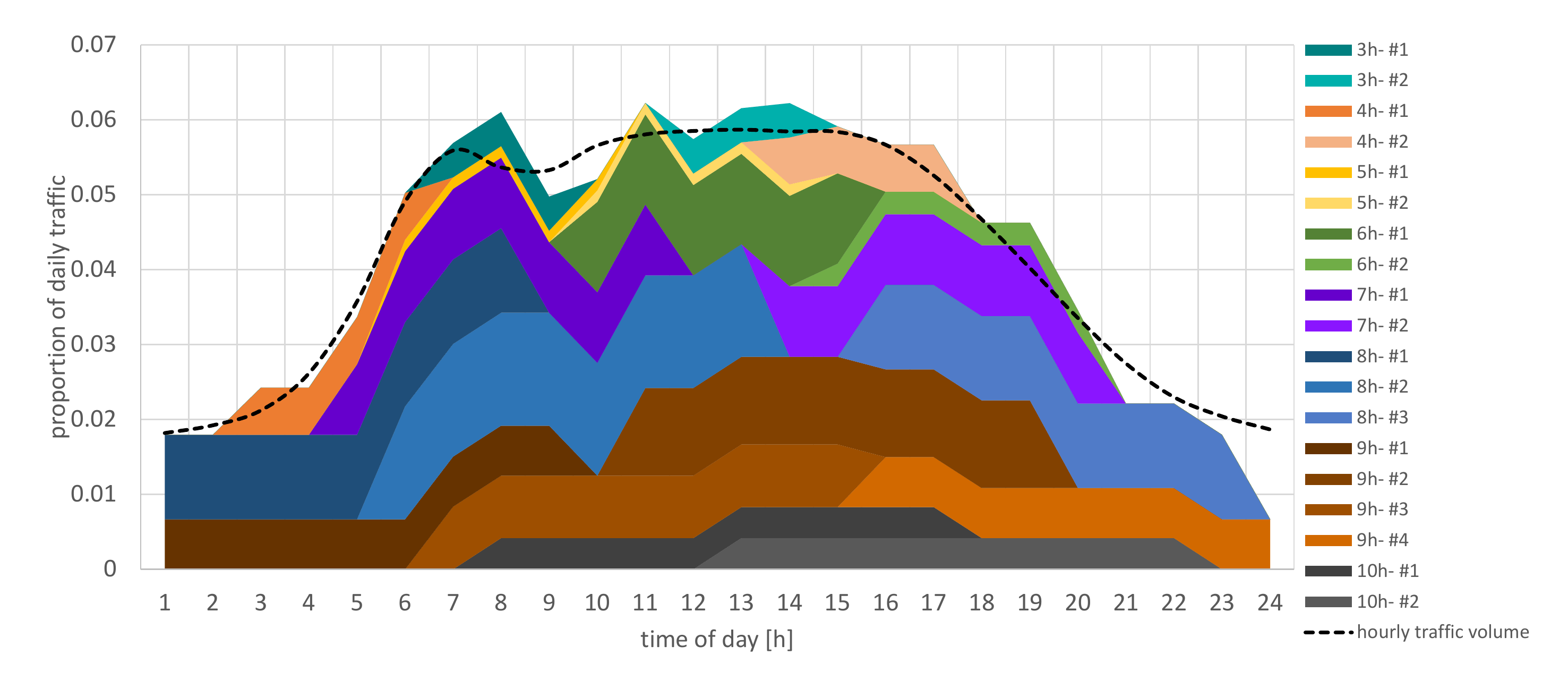}
    \caption{Distribution of daily driving patterns over the course of the day}
    \label{fig_si05}
\end{figure}

The traffic model data does not include information on the starting time of a driving profile. Therefore, in the next step, the profiles are distributed over the day according to the empirically observed course of the daily mileage \cite{bast_2023}, taking into account their relative frequency (i.e. the mileage represented by them). For each daily operating period, two different starting times are defined. The operating durations of eight and nine hours, which are particularly strongly represented, are given three and four different start times respectively. The resulting temporal distribution of traffic volumes is shown in Figure~\ref{fig_si05}. The contribution of the example profile shown above (eight hours operating time, with three different starting times) corresponds to the dark blue coloured areas in Figure~\ref{fig_si05}. The resulting electricity demands and charging availabilities of the profiles are fed into the DIETER model, and are available in the public repository \url{https://gitlab.com/diw-evu/projects/my_eroads}.

With the procedure described, we have generated an ensemble of operational profiles for domestic German heavy road freight transport that reflect well both the macroscopic parameters of the German transport sector (mileage, distribution of daily driving distances, daily pattern of traffic volume) and operational boundary conditions (loading / unloading times, mandatory driver’s breaks). 

\subsubsection{Fuel demand profiles of FCEV and Ptl trucks}

For FCEV and PtL trucks, no differentiated driving and charging availability time series are needed; instead, only aggregate fuel demand time series are required. These are based on the same empirical traffic data from German highways \cite{bast_2023}, which are also used for deriving the (ERS-)BEV profiles. For FCEV, we assume a specific hydrogen demand of 6.8~kg/100km, and for ICEV-PtL a specific Diesel demand of 27.1~l/100km. Assuming the overall HDV fleet to be either FCEV or ICEV-PtL, this corresponds to a yearly hydrogen demand of around 68~TWh, or a yearly e-fuel demand of 80~TWh, respectively. The resulting fuel demand time series are available together with all other input parameters in the public repository \url{https://gitlab.com/diw-evu/projects/my_eroads}.

\clearpage

\subsection{Supplemental Notes}


\subsubsection{Additional results for baseline model: Electricity storage capacity and use}\label{sec: si storage}

In the following, we provide additional results for the baseline model specification discussed in the main part of the paper. This includes the effects of electrified HDV on capacity and yearly use of electricity storage, electrolysis capacity, the shares of renewable energy sources in overall demand, and carbon emissions, as well as additional time series results.

\begin{figure}[ht!]
    \begin{mdframed}
        \centering
        \includegraphics[width =0.65\textwidth]{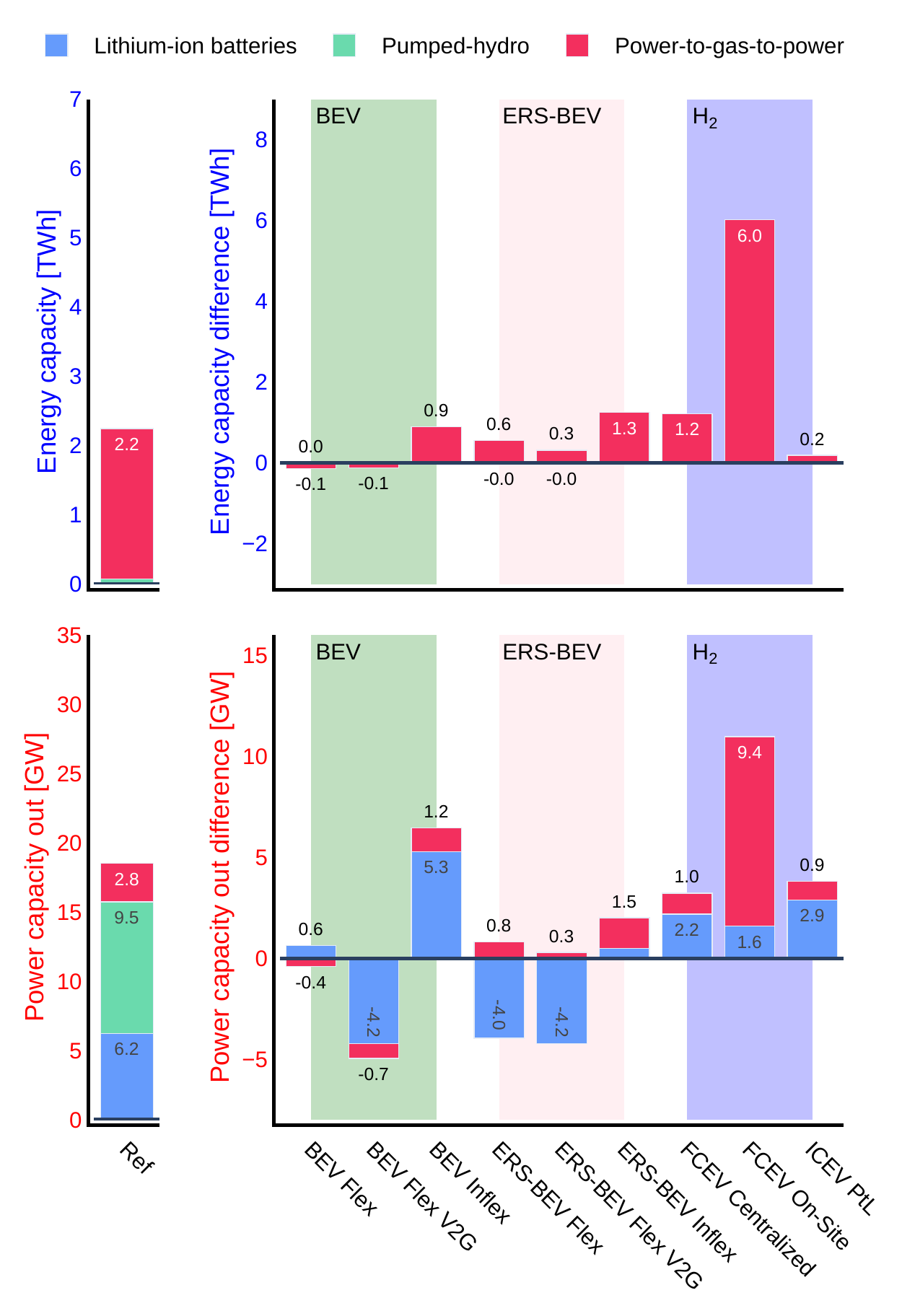}
    \end{mdframed}
    \caption{Effects of different HDV options on optimal electricity storage energy capacity (top panel), and on storage output power capacity (lower panel)}
    \label{fig_si06}
\end{figure}

Figure~\ref{fig_si06} shows optimal electricity storage energy capacities in the reference and the changes induced by HDV (upper panel), as well as optimal storage output power rating in the reference, and corresponding effects of HDV (lower panel). Effects are differentiated by technology, i.e.,~for lithium-ion batteries and power-to-gas-to-power storage. Batteries come with high energy-specific investment costs and low power-specific costs, which makes them a typical short-duration storage technology with a low energy-to-power ratio and many yearly cycles; power-to-gas-to-power, conversely, comes with very low energy-specific costs and higher power-specific costs, and is thus optimally used as a long-duration storage technology \cite{sepulveda_2021}. Pumped hydro storage is assumed to be fixed to TYNDP scenario assumptions because of limited expansion opportunities in Germany.

As for storage energy capacity, long-duration electricity storage clearly dominates because of much lower energy-specific investment costs. We further find the largest impact of HDV on electricity storage in the case of on-site, i.e.,~inflexible, hydrogen provision. Here, the installed energy capacity is nearly four times as high as in the reference. As hydrogen has to be generated largely at the time of demand, and the potential to expand fossil generators is limited, this means that long-duration storage is used to balance variable renewable electricity generation with electricity demand for hydrogen production. For inflexible BEV or ERS-BEV, as well as centralized hydrogen supply, long-duration electricity storage capacity is expanded to a much smaller extent; and it is nearly zero for centralized PtL despite the highest additional electricity demand, as liquid e-fuels come with even cheaper bulk storage options.

When it comes to optimal storage output power, (in-)directly electrified HDV have a noticeable effect on both batteries and long-duration storage technologies (lower panel of Figure~\ref{fig_si06}). Overall, BEV or ERS-BEV with V2G substantially reduce the required storage power, as truck batteries substitute stationary batteries, which would otherwise be needed to balance diurnal fluctuations of solar PV. Inflexible charging of BEV or ERS-BEV conversely increases optimal storage power. For long-duration storage, the impacts of HDV on power output capacity are largely similar to those for energy capacity, with by far the largest effect for inflexible on-site hydrogen generation.

\subsubsection{Additional results for baseline model: Electrolysis capacity}\label{sec: electrolysis capacity}

\begin{figure}[ht!]
    \begin{mdframed}
        \centering
        \includegraphics[width =0.27\textwidth]{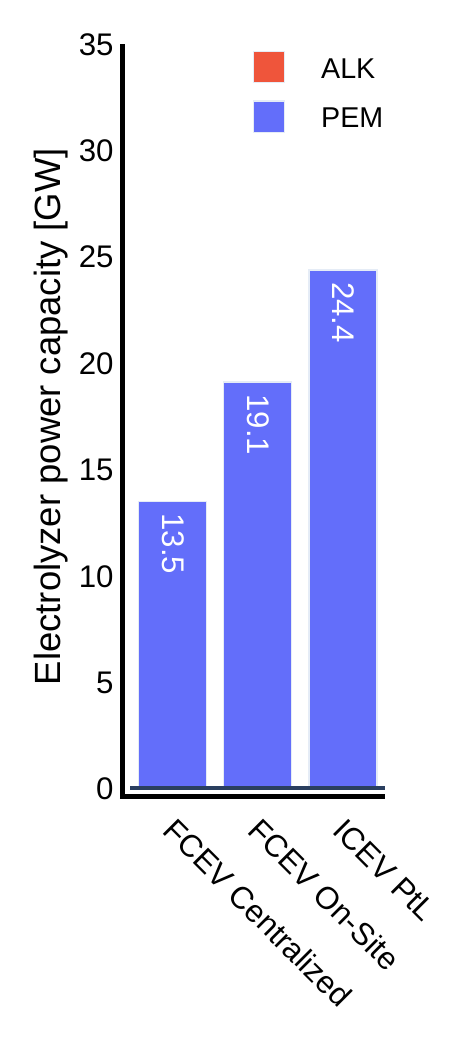}
    \end{mdframed}
    \caption{Optimal electrolysis capacity}
    \label{fig_si07}
\end{figure}

Optimal investments into electrolyzers are shown in Figure~\ref{fig_si07}. Note that hydrogen and e-fuels are assumed to be produced in Germany via electrolysis and, in the case of PtL, using a non-specified but low-cost carbon source. Only investments into proton exchange membrane (PEM) electrolyzers are made, which have lower conversion losses, but higher specific investment costs than alkaline electrolyzers (ALK). This is because renewable energy is scarce under baseline model assumptions, and higher energy efficiency of electrolyzers accordingly matters more than lower investment costs. This changes if Germany is modeled in isolation, as this leads to larger renewable surplus generation, which in turn makes lower-cost ALK technology a part of the optimal electrolysis technology mix (compare also~\cite{stoeckl_2021}). 

The electrolysis capacity required to supply the German HDV fleet is highest in the PtL scenario with 24.4~GW, as e-fuels require more hydrogen than the two FCEV cases. Among the hydrogen scenarios, inflexible on-site generation further leads to a higher optimal electrolysis capacity of 19.1~GW compared to 13.5~GW for centralized electrolysis. This is, again, because on-site electrolyzers have to produce hydrogen with a similar time profile as the hydrogen demand at filling stations. In contrast, centralized electrolyzers can use low-cost production-site hydrogen storage to balance fluctuations of renewable supply and hydrogen demand.

\subsubsection{Additional results for baseline model: Renewable shares of the German power sector}\label{sec: renewable shares}

\begin{figure}[ht!]
    \begin{mdframed}
        \centering
        \includegraphics[width =0.55\textwidth]{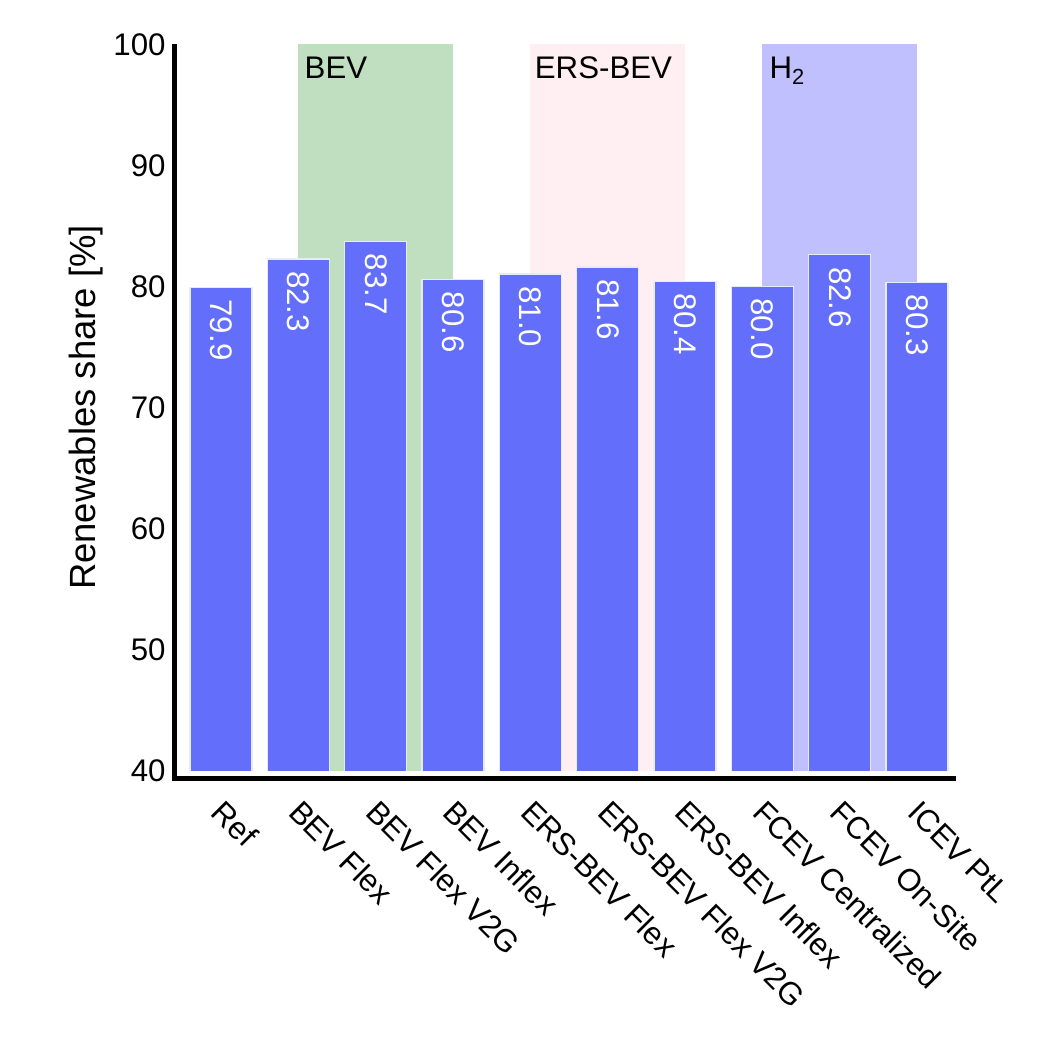}
    \end{mdframed}
    \caption{Share of renewable electricity generation in total electricity demand for Germany}
    \label{fig_si08}
\end{figure}

Figure~\ref{fig_si08} shows the shares of renewable electricity generation in electricity demand for Germany for alternative HDV electrification scenarios. Flexible BEV and ERS-BEV, especially if combined with V2G, increase the renewable share by almost up to four percentage points in the BEV Flex V2G scenario. This is because HDV batteries can foster the integration of additional solar PV generation, even beyond the electricity demand related to the vehicles themselves. This is, in contrast, not the case for inflexibly charged BEV or ERS-BEV.

Centralized hydrogen and PtL supply chains hardly increase the optimal renewable share, despite high flexibility potentials related to their low-cost centralized storage options. This is because these flexibility benefits are outweighed by the sheer amount of additional renewable electricity needed for HDV, which is substantially higher than in BEV or ERS-BEV cases (compare Figure~\ref{fig_02}) and thus causes increasing renewable integration efforts. On-site electrolysis, in turn, leads to a higher increase in the renewable share because its temporal inflexibility induces higher long-duration electricity storage investments that allow for better integration of variable renewables.

\subsubsection{Additional results for baseline model: Direct carbon emissions}\label{sec: CO2}

\begin{figure}[ht!]
    \begin{mdframed}
        \centering
        \includegraphics[width =0.99\textwidth]{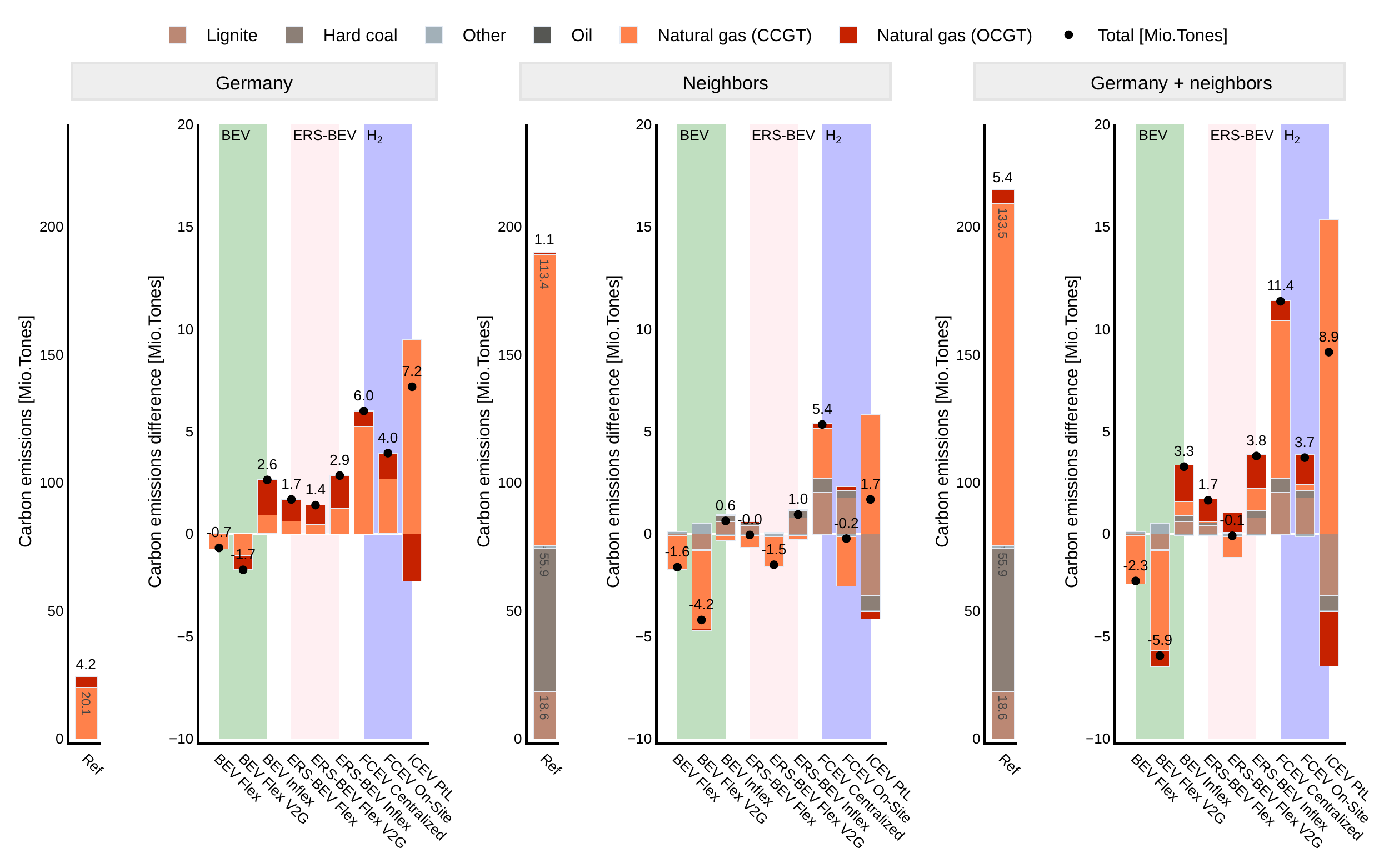}
    \end{mdframed}
  \caption{Direct CO$_{2}$ emissions from electricity generation. The chart displays emissions from Germany, its Neighbors and the two together. Every group contains two subplots; the subplot on the left shows the overall emissions of the reference without electrified HDV, and the right-hand side subplot shows differences between HDV scenarios and the reference.}
  \label{fig_si09}
\end{figure}

Among all scenarios, CO$_{2}$ emissions increase the most if the HDV fleet uses e-fuels or hydrogen (Figure~\ref{fig_si09}). In these scenarios, this is a direct consequence of additional electricity generation from natural gas, and to a smaller extent from lignite and hard coal in neighboring countries. Among the two hydrogen cases, on-site electrolysis at filling stations leads to lower emission impacts, as its temporal flexibility restrictions in combination with the assumed transmission capacities limit the possibility of increasing imports or generation from natural gas plants; instead, additional renewable energy, combined with long-duration electricity storage, is used (compare section~\ref{sec: capacity and dispatch effects}). Emission effects are smaller for BEV and ERS-BEV, and even negative for flexibly charged BEV, especially if combined with V2G. The latter is driven by an additional expansion of solar PV facilitated by V2G. For neighboring countries, relative emission effects are smaller, as by assumption, they have lower renewable energy shares and, in turn, higher emissions, as well as no electrified truck fleets.

\subsubsection{Additional results for baseline model: Electricity time series}\label{sec: si time series}

\begin{figure}[ht!]
    \begin{mdframed}
        \centering
        \includegraphics[width =0.96\textwidth]{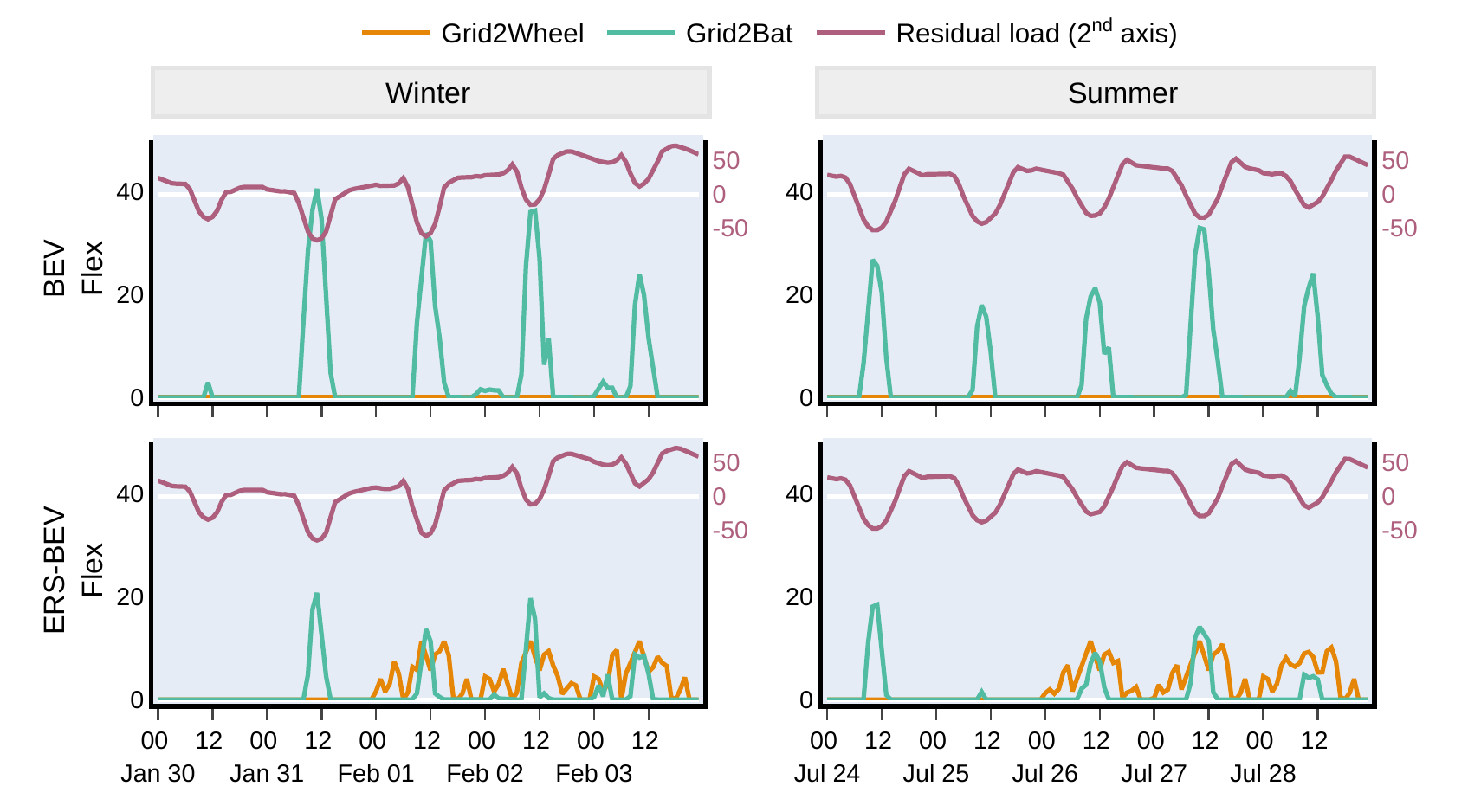}
    \end{mdframed}
    \caption{Additional time series of electricity demand, complementary to Figure~\ref{fig_04} (unit: GW). The chart shows a sample of five exemplary days starting with Saturday.}
    \label{fig_si10}
\end{figure}

Figure~\ref{fig_si10} shows additional electricity time series for the BEV~Flex and ERS-BEV~Flex scenarios without V2G, which are omitted in Figure~\ref{fig_04} because of space restrictions. The Figure also differentiates the grid electricity demand of HDV for electricity directly used for driving in ERS-BEV (Grid2Wheel) and battery charging (Grid2Bat). Flexible BEV, similar to BEV Flex V2G, use renewable surplus energy (negative residual load) for charging. Flexible ERS-BEV effectively use overhead line electrification while driving in combination with optimal charging of batteries at charging stations.

\begin{figure}[htp!]
    \begin{mdframed}
        \centering
        \includegraphics[width =0.96\textwidth]{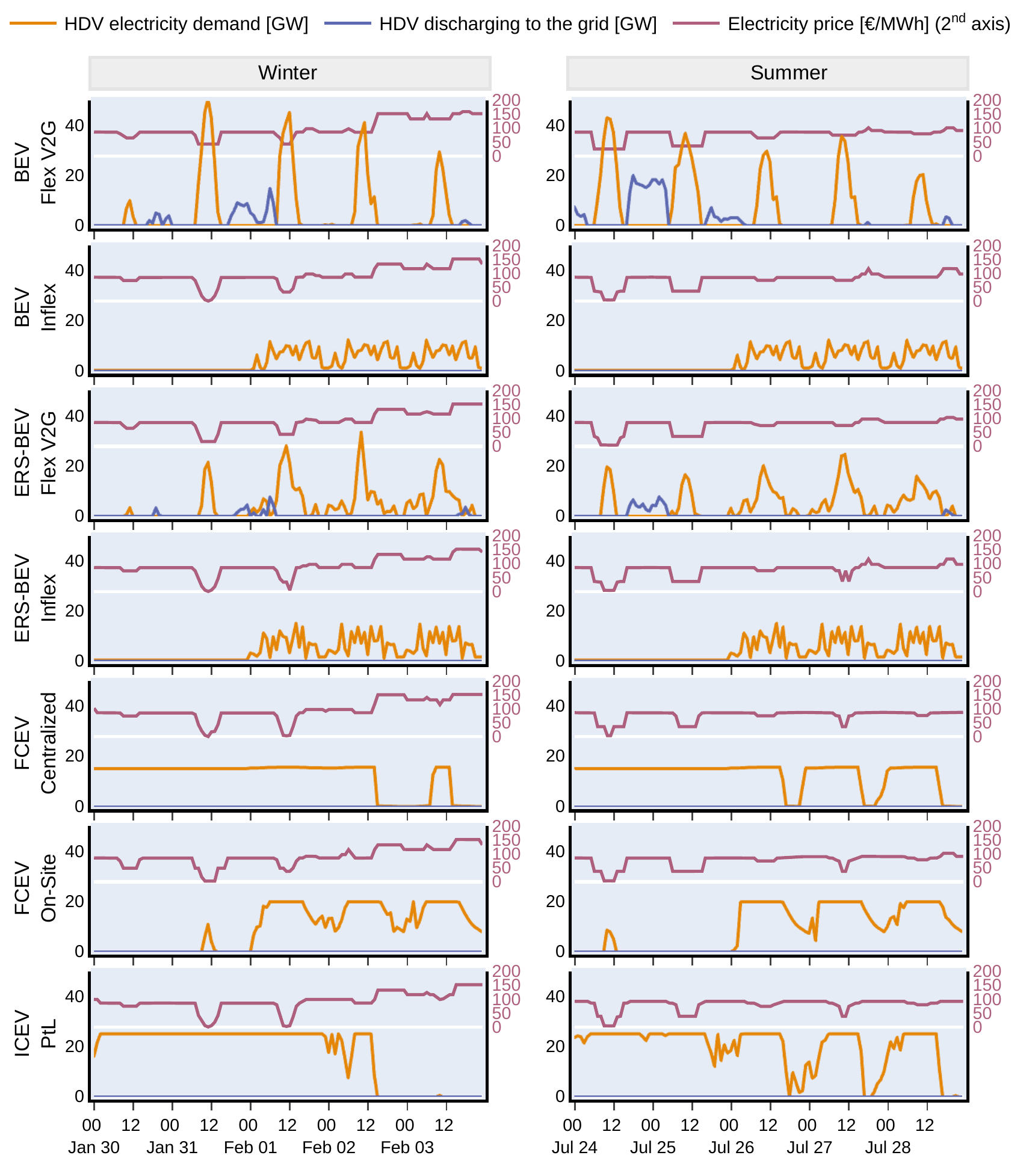}
    \end{mdframed}
    \caption{Sample of electricity time series five exemplary days: ``HDV electricity demand'' illustrates electricity flowing to BEV or ERS-BEV batteries, including energy later discharged back via V2G, as well as electricity needs of hydrogen and PtL supply chains; ``HDV discharging to the grid'' are V2G power flows from BEV or ERS-BEV to the grid; and ``Electricity price'' is the wholesale electricity market price in the respective hour. The samples start on a Saturday.}
    \label{fig_si11}
\end{figure}

Figure~\ref{fig_si11} is complementary to Figure~\ref{fig_04} as it does not show residual load, but wholesale electricity market prices. These largely follow the residual load; prices are generally lower when the residual load is lower.

\begin{figure}[htp!]
    \begin{mdframed}
        \centering
        \includegraphics[width =0.96\textwidth]{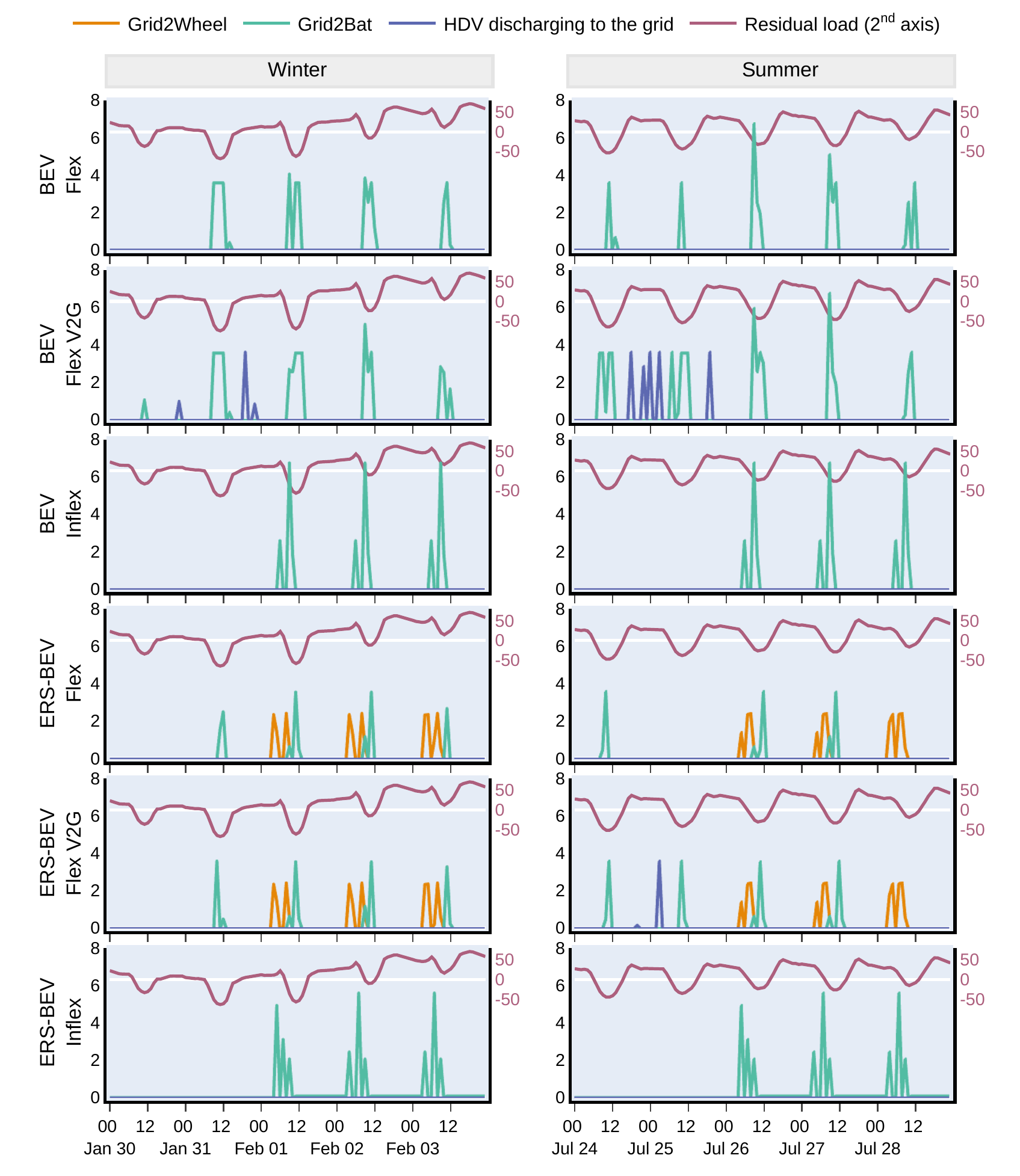}
    \end{mdframed}
    \caption{Time series of electricity demand for vehicle profile nr. 05 (unit: GW)}
    \label{fig_si12}
\end{figure}

\begin{figure}[htp!]
    \begin{mdframed}
        \centering
        \includegraphics[width =0.96\textwidth]{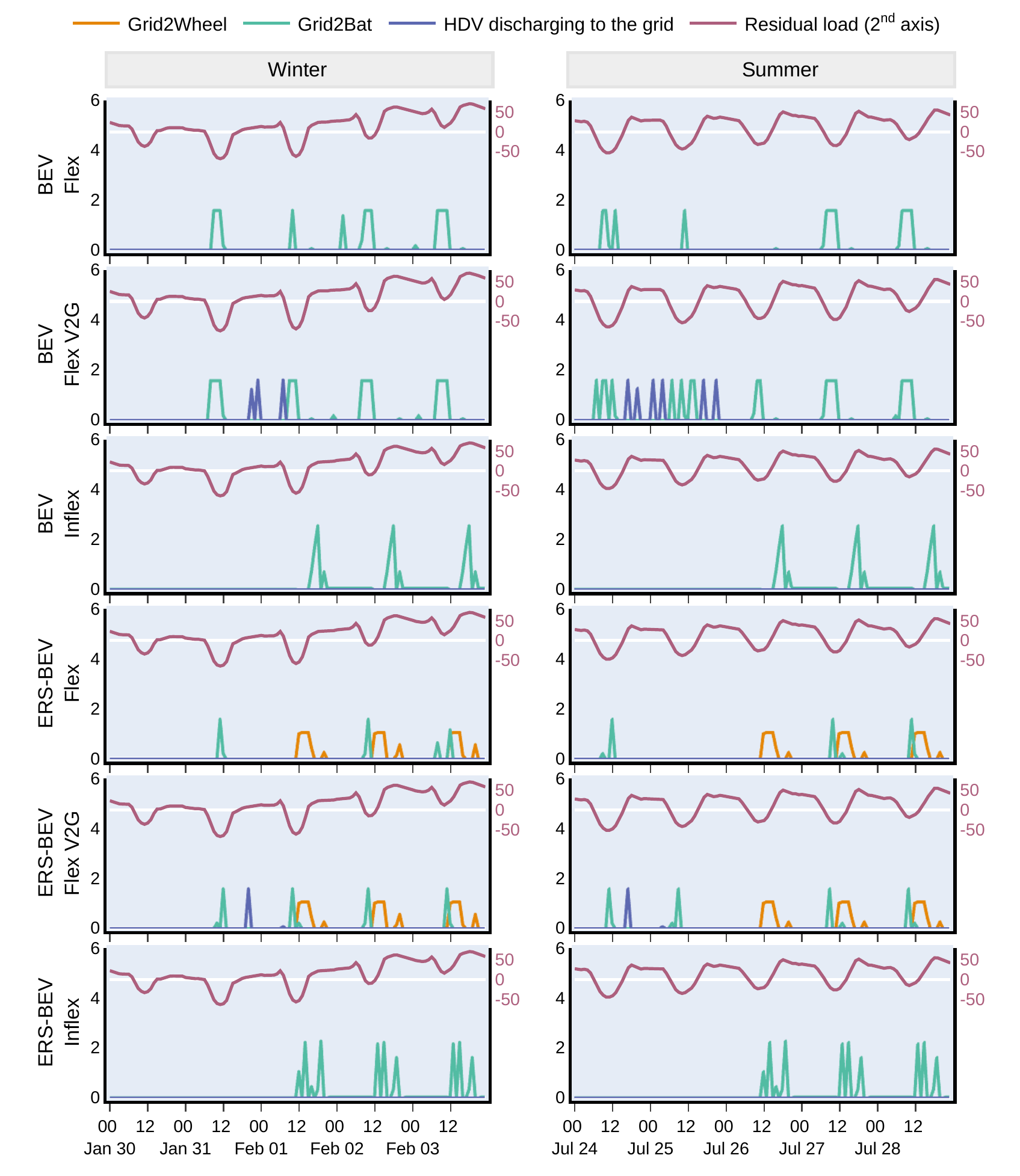}
    \end{mdframed}
    \caption{Time series of electricity demand for vehicle profile nr. 16 (unit: GW)}
    \label{fig_si13}
\end{figure}

Figures~\ref{fig_si12} and \ref{fig_si13} show additional exemplary time series not for the aggregate HDV fleet, but for two specific vehicle profiles. Vehicle profile~5 represents a truck with a driving pattern in the morning between 4~am and 11~am, and profile~16 is driving between 12~noon and 9~pm with one hour stop. Both vehicles have a charging station available before and after every trip. These profiles are ``peakier'' than aggregate fleet profiles. The sequence of Grid2Bat and V2G operations is more clearly visible than in the aggregate fleet time series. It can also be seen that ERS-BEV are partly charging their batteries while connected to the overhead lines, which is relevant for the V2G configuration, especially when the residual load is negative.

\clearpage


\subsubsection{Sensitivity analyses: Effects of the European interconnection}\label{sec: sensitivity island}

\begin{figure}[ht!]
    \begin{mdframed}
        \centering
        \includegraphics[width =\textwidth]{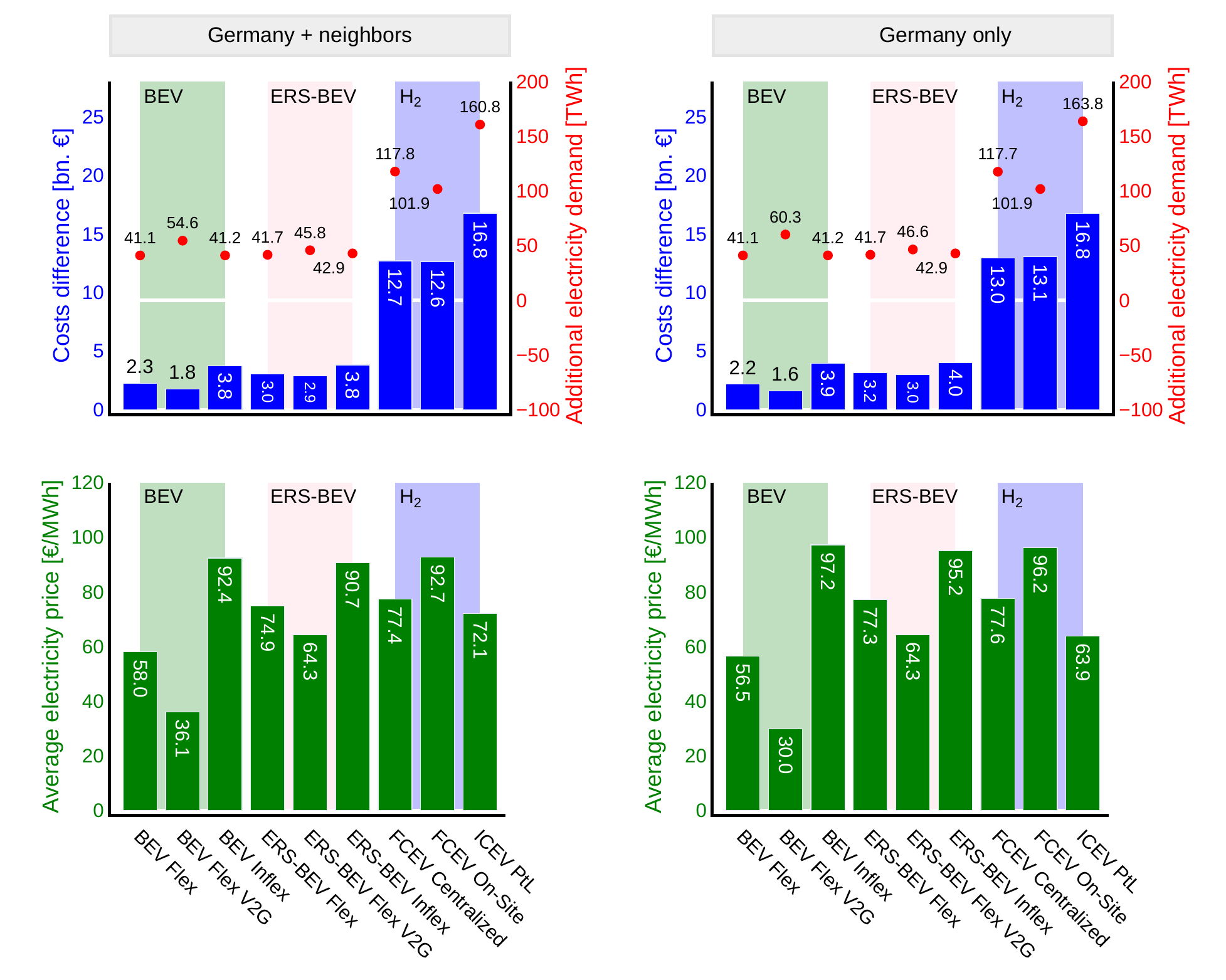}
    \end{mdframed}
    \caption{Changes in yearly power sector costs and electricity demand induced by different HDV options (top panels) and average prices of charging electricity (lower panels), evaluating the effects of the European interconnection}
    \label{fig_si14}
\end{figure}

In the main part of the paper, we show results for a central European interconnection where the integration of variable renewables and electrified HDV in Germany benefits from the flexibility provided by geographical balancing. It has been shown that the flexibility provided by the European interconnection reduces electricity storage needs, as it allows smoothing complementary wind power profiles in different countries \cite{roth_2023}. In the following, we compare results for alternative model runs where the German power sector is modeled in isolation. This setting is less realistic, but helps to separate the effects of the European interconnection on results. It also gives a qualitative indication of how results are potentially distorted in model analyses that focus on single countries.

Results are qualitatively similar for a case where the German power sector is modeled in isolation from its neighboring countries, but slightly more pronounced (top right panel of Figure~\ref{fig_si14}). Here, flexible BEV with V2G are even more beneficial than in the setting where Germany is interconnected with its neighboring countries (1.6~bn Euros vs.~1.8~bn Euros additional costs). This is because in such a flexibility-constrained setting, electricity storage provided by HDV batteries becomes even more beneficial for integrating variable renewables. On the contrary, inflexible BEV operations lead to a slightly higher cost increase (3.9~bn Euros vs.~3.8~bn Euros) than in a case with interconnection.

If Germany is modeled as an electric island, average electricity prices for HDV options that are particularly flexible (BEV~Flex, BEV~Flex~V2G, ICEV~PtL) are lower than in the interconnected case (lower right panel of Figure~\ref{fig_si14}). The reason is that flexible HDV can make use of more pronounced renewable surplus generation events which go along with lower prices. In contrast, temporally less flexible options are forced to charge their batteries in hours with higher average prices.

\begin{figure}[ht!]
    \begin{mdframed}
        \centering
        \includegraphics[width =\textwidth]{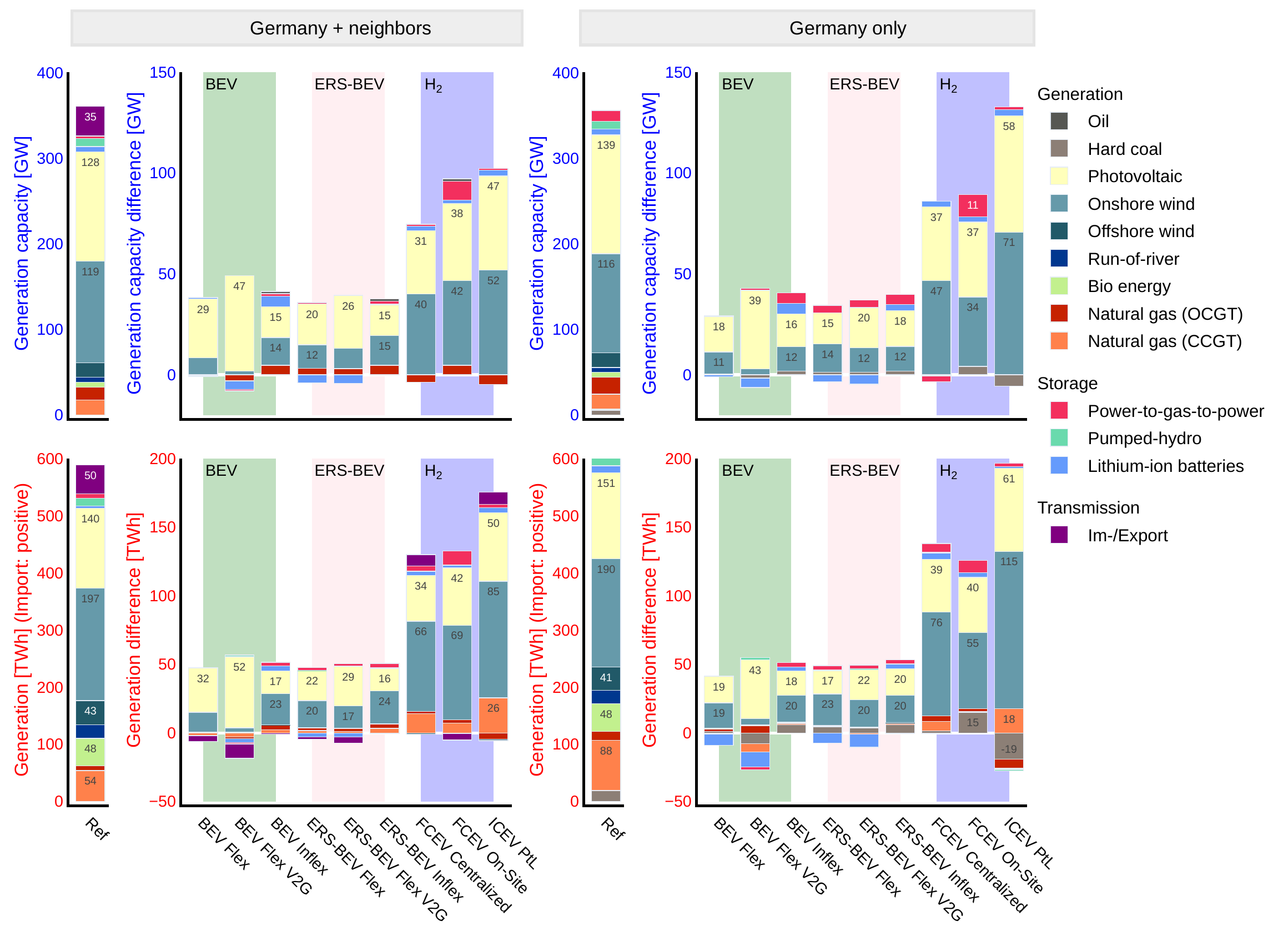}
    \end{mdframed}
    \caption{Effects of different HDV options on optimal electricity generation capacity (top panels), and on yearly electricity generation (lower panels), evaluating the effects of the European interconnection}
    \label{fig_si15}
\end{figure}

When Germany's electricity sector is considered in isolation, the missing European interconnection no longer allows for import capacity, so both overall and firm capacity needs to increase (top right panels of Figure~\ref{fig_si15}). Further, HDV-induced effects on optimal generation capacities are generally smaller than if the interconnection with neighboring countries is considered. This is because the reference (Ref) of an isolated setting already has a larger power plant fleet, with more under-utilized capacity available. This particularly benefits the temporally flexible BEV and ERS-BEV options (especially with V2G), as these allow making use of parts of the renewable surplus energy present in the reference. In the hydrogen and PtL cases, this effect vanishes or even reverses, as the additional load induced by HDV is much higher than the renewable surplus energy available in the reference.

\begin{figure}[ht!]
    \begin{mdframed}
        \centering
        \includegraphics[width =\textwidth]{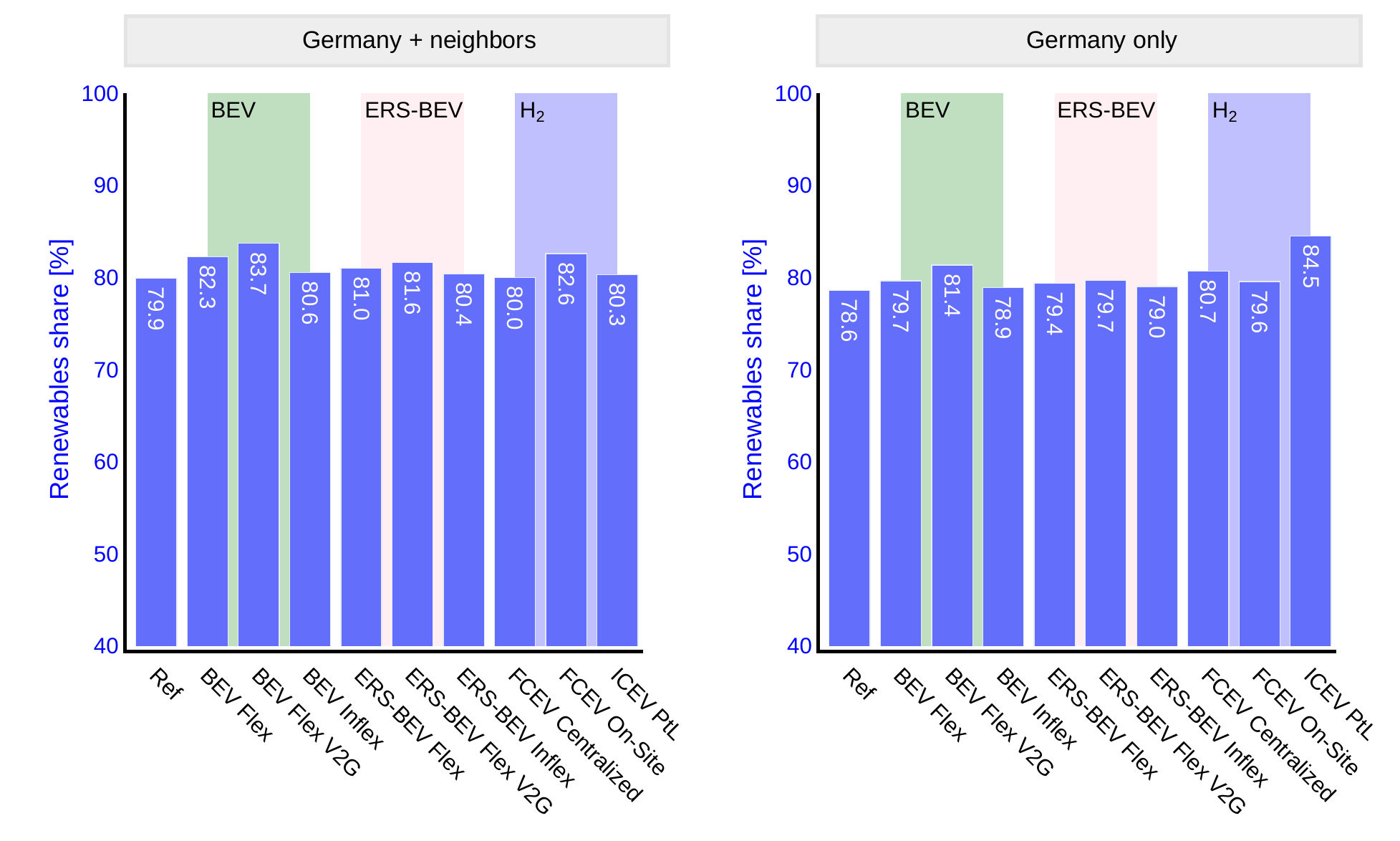}
    \end{mdframed}
    \caption{Renewable share of the German power sector considering scenarios with neighboring countries (left panel), and with Germany modeled in isolation (right panel)}
    \label{fig_si16}
\end{figure}

The resulting renewable share of the German power sector tends to be lower if Germany is modeled as an electric island (Figure~\ref{fig_si16}). This is again because the missing flexibility of the European interconnection makes it harder to integrate variable renewables. The ICEV~PtL case is an exception. Here, low-cost e-fuel storage enables higher use of solar PV and wind power, and hard coal is no longer used for electricity generation.

\clearpage

\subsubsection{Sensitivity analyses: Effects of a capacity constraint for wind power}\label{sec: sensitivity wind}

In the central part of the analysis, we assume that solar PV and wind power investments are possible without any limits. We consider this to be a meaningful setting for highlighting general effects, which should also be generally applicable for other geographic settings. Yet in the specific German case, the medium-term capacity expansion of onshore and offshore wind power may, in fact, be constrained. Considering long lead times related to planning and admission processes, we specify a less optimistic scenario where we assume that onshore and offshore wind power expansion may not exceed 100~GW and 30~GW by 2030, respectively. 

\begin{figure}[ht!]
    \begin{mdframed}
        \centering
        \includegraphics[width =\textwidth]{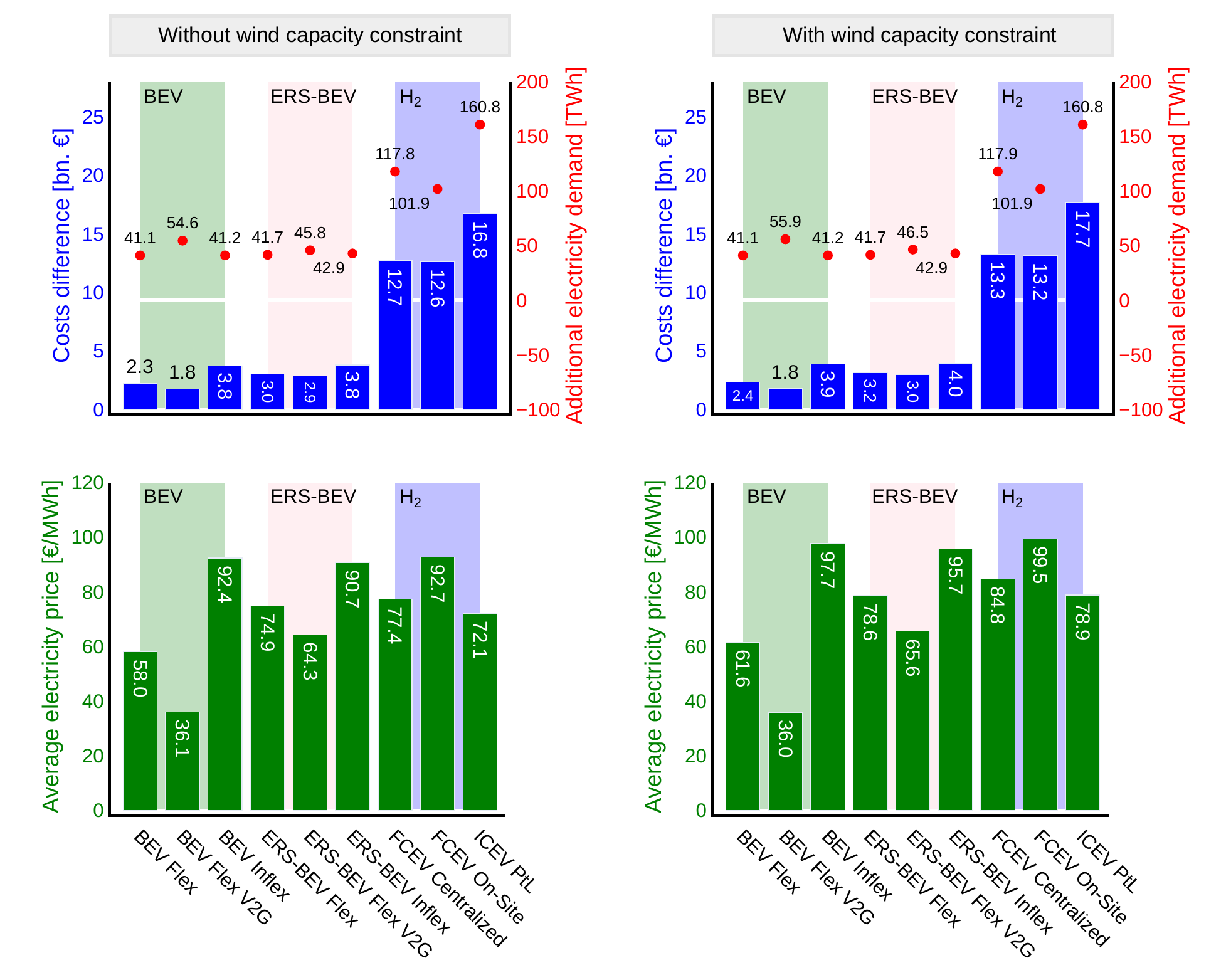}
    \end{mdframed}
    \caption{Changes in yearly power sector costs and electricity demand induced by different HDV options (top panels), and average prices of charging electricity (lower panels), evaluating the effects of capacity constraints for wind power}
    \label{fig_si17}
\end{figure}

\begin{figure}[ht!]
    \begin{mdframed}
        \centering
        \includegraphics[width =\textwidth]{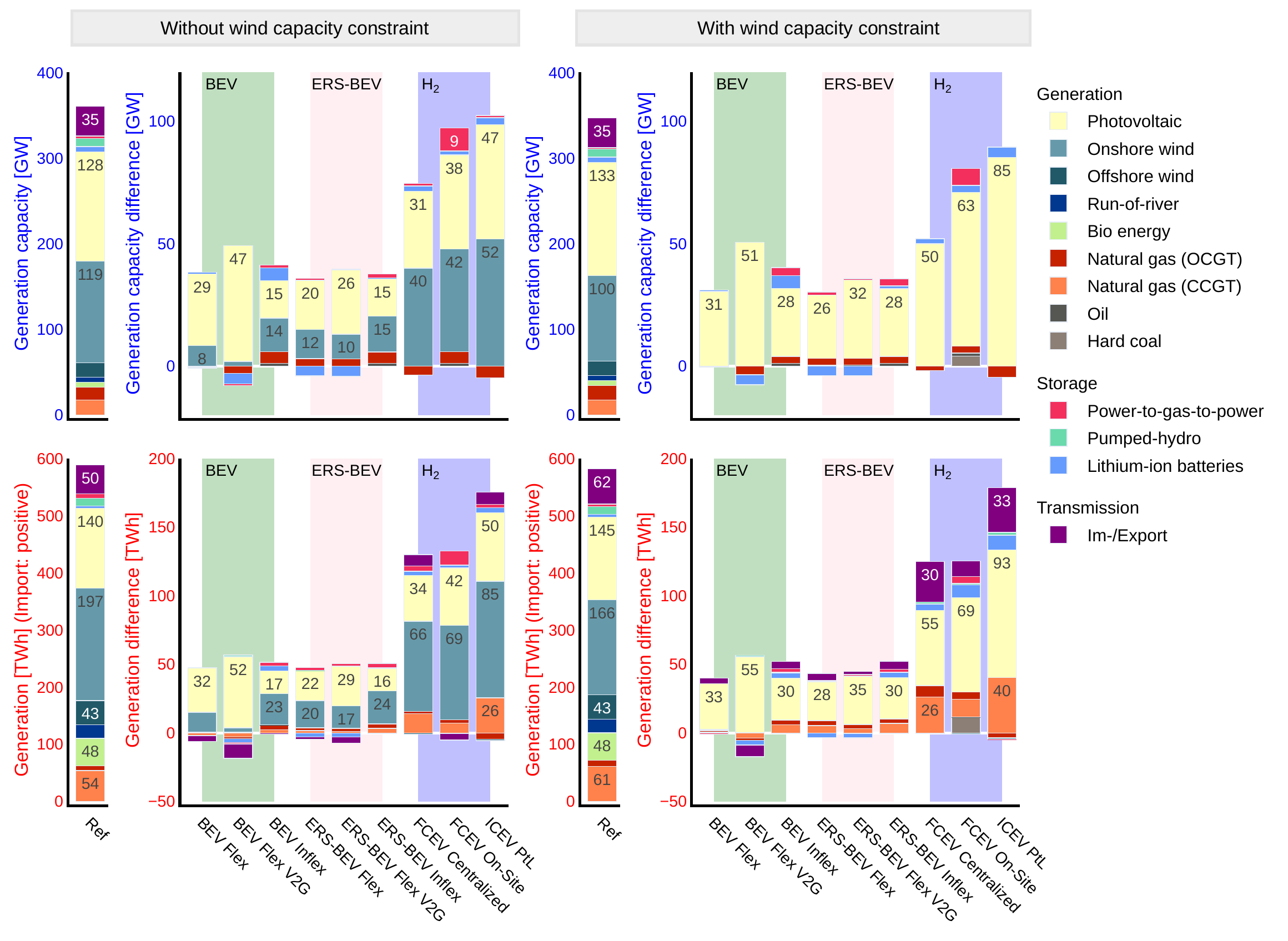}
    \end{mdframed}
    \caption{Effects of different HDV options on optimal electricity generation capacity (top panels), and on yearly electricity generation (lower panels), evaluating the effects of capacity constraints for wind power}
    \label{fig_si18}
\end{figure}

Figure~\ref{fig_si17} again shows cost differences of various HDV options against a reference without HDV (top panels). As expected, the case with wind capacity constraints on the right-hand side leads to slightly higher additional costs. The lower expansion of wind capacity is compensated by increasing solar PV investments, as shown in Figure~\ref{fig_si18} (top right panels), which are slightly more expensive in Germany. Only in the case of flexible BEV with V2G, costs and average charging electricity prices do not increase compared to baseline assumptions. This is because the short-duration electricity storage characteristics of V2G fits particularly well to solar PV generation profiles, which also helps avoiding electricity generation from natural gas and substituting firm capacity to a certain degree. 

\begin{figure}[ht!]
    \begin{mdframed}
        \centering
        \includegraphics[width =\textwidth]{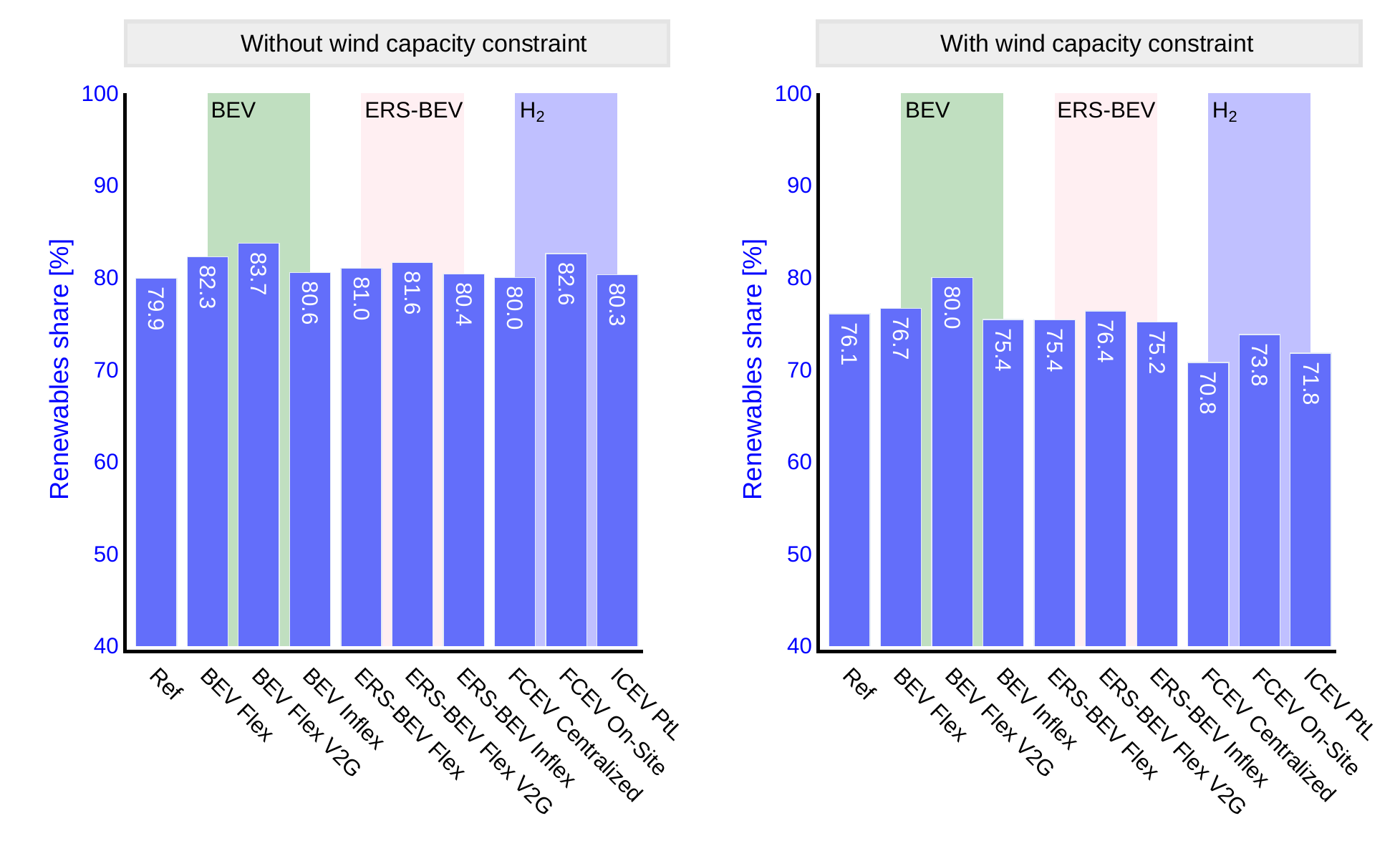}
    \end{mdframed}
    \caption{Renewable share of the German power sector considering scenarios without (left panel) and with wind capacity constraints (right panel)}
    \label{fig_si19}
\end{figure}

Yearly electricity generation results (Figure~\ref{fig_si18}, lower panels) shows that the German power sector relies more on imports and natural gas when wind capacity is constrained. This is particularly significant for FCEV and hydrogen-based PtL. That is, wind power is only partly replaced by solar PV, and more fossil-fueled power generation is used in almost all HDV configurations. Accordingly, the renewable share of a power sector with constrained wind power expansion is also lower than under baseline assumptions (Figure~\ref{fig_si19}).

\subsubsection{Sensitivity analyses: Effects of alternative assumptions on charging availability}\label{sec: sensitivity charging availability}

\begin{figure}[ht!]
    \begin{mdframed}
        \centering
        \includegraphics[width =\textwidth]{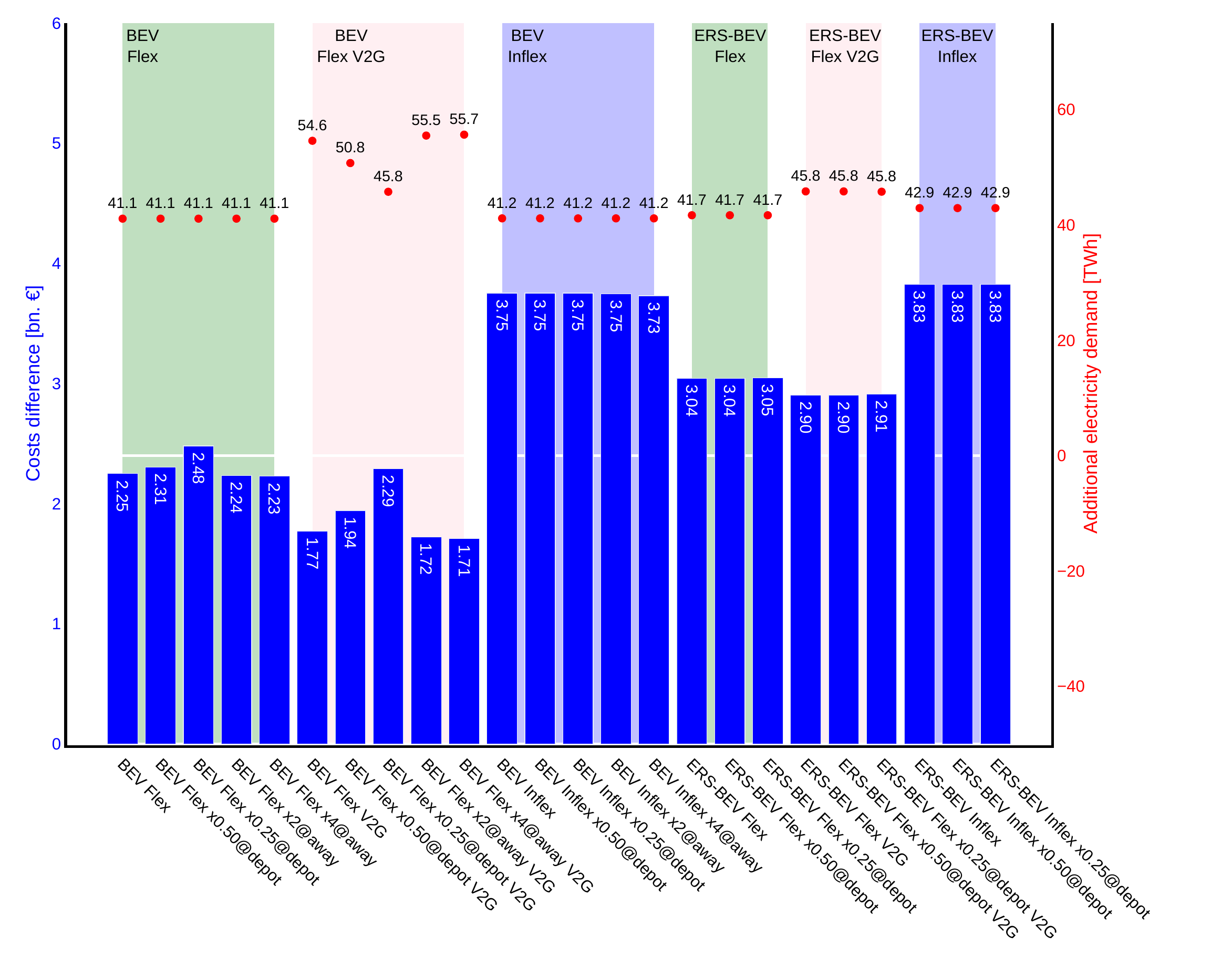}
    \end{mdframed}
    \caption{Changes in yearly power sector costs and electricity demand induced by different HDV options, including the effects of alternative assumptions on charging availability. ``@depot'' refers to lower charging availability at depots, ``@away'' to higher charging availability at loading/unloading stops and during mandatory driver's breaks.}
    \label{fig_si20}
\end{figure}

\begin{figure}[ht!]
    \begin{mdframed}
        \centering
        \includegraphics[width =\textwidth]{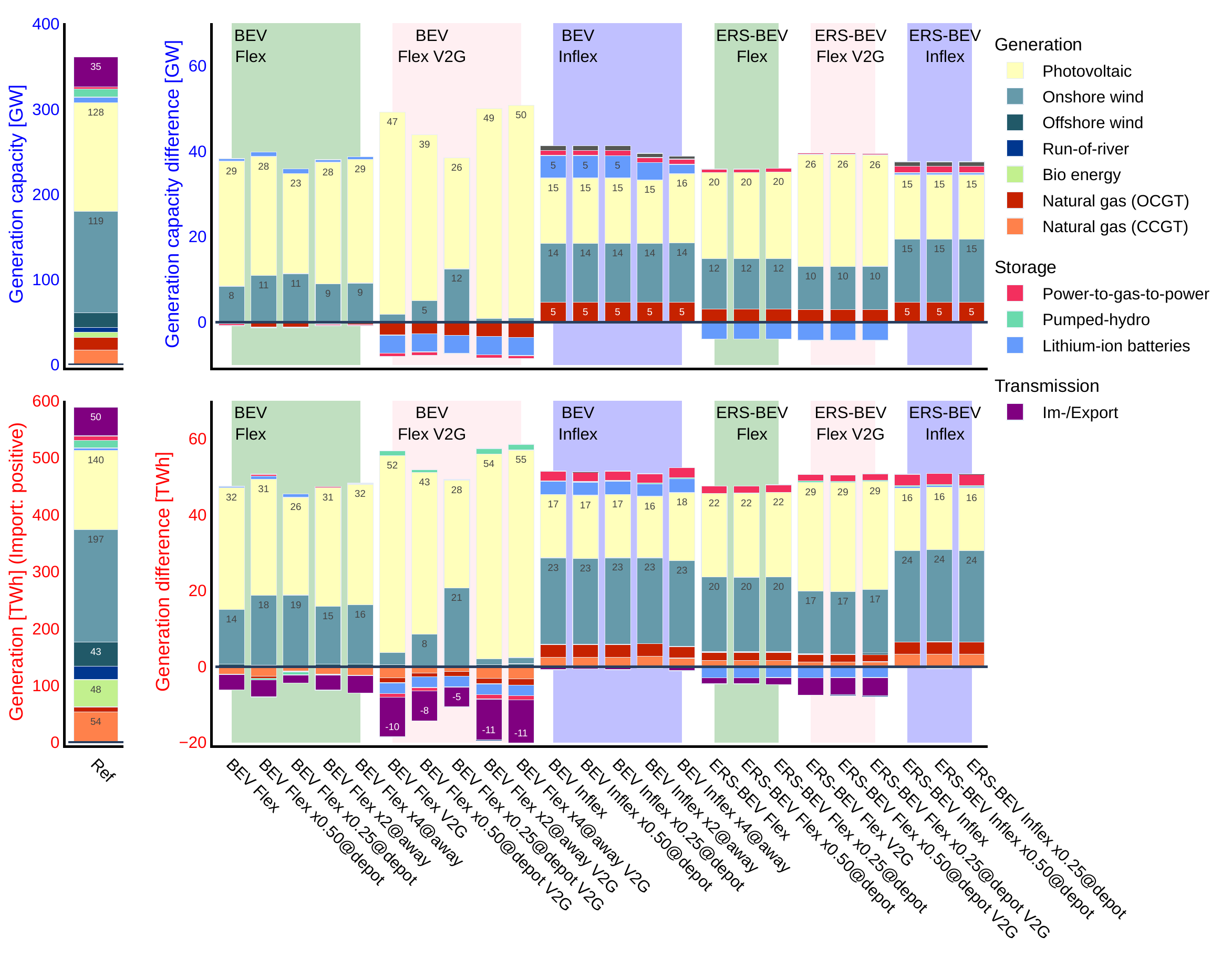}
    \end{mdframed}
    \caption{Effects of different HDV options on optimal electricity generation capacity (top panels), and on yearly electricity generation (lower panels), including the effects of alternative assumptions on charging availability. ``@depot'' refers to lower charging availability at depots, ``@away'' to higher charging availability at loading/unloading stops and during mandatory driver's breaks.}
    \label{fig_si21}
\end{figure}

\begin{figure}[ht!]
    \begin{mdframed}
        \centering
        \includegraphics[width =\textwidth]{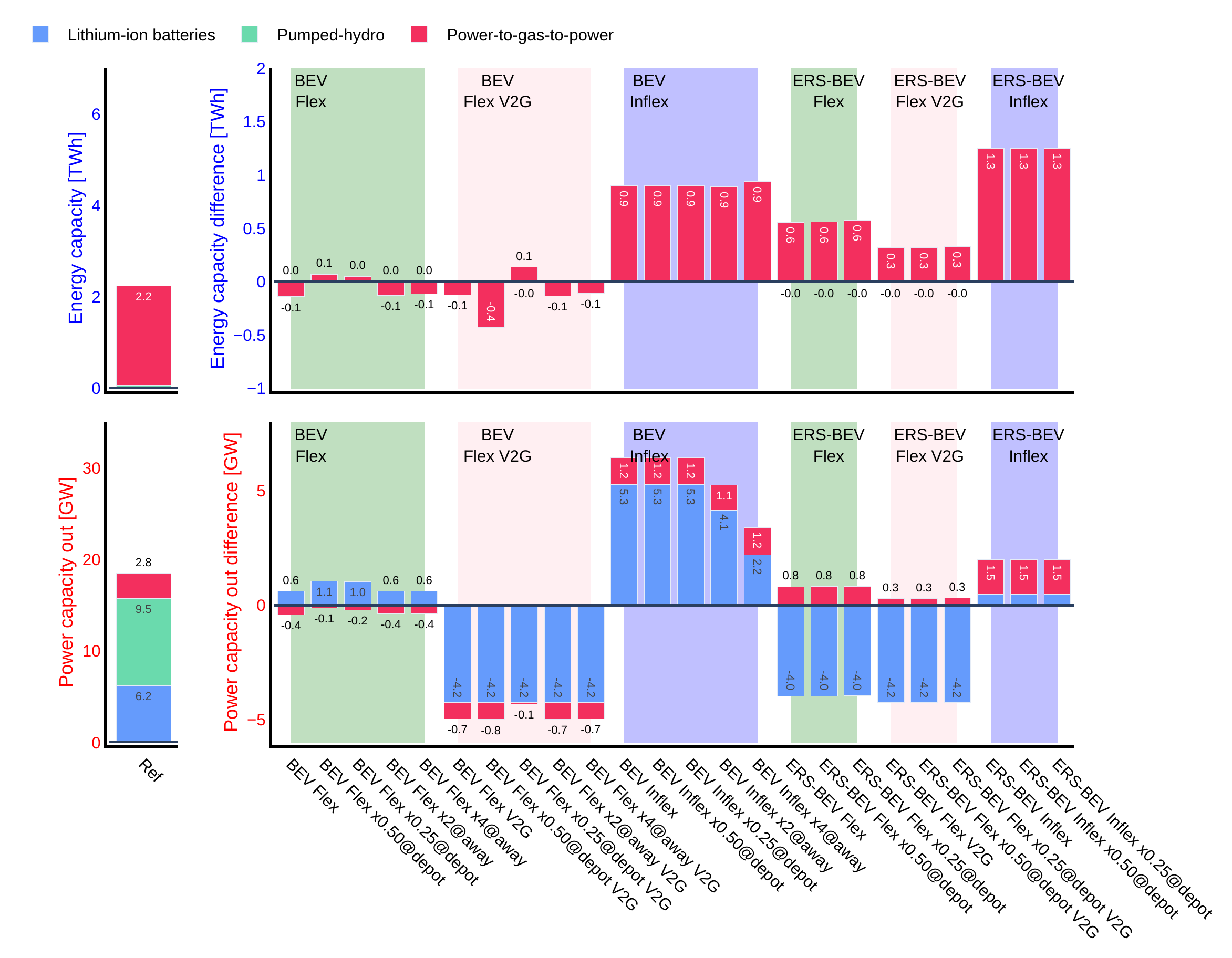}
    \end{mdframed}
    \caption{Effects of different HDV options on optimal electricity storage energy capacity (top panel), and on storage output power capacity (lower panel), including the effects of alternative assumptions on charging availability. ``@depot'' refers to lower charging availability at depots, ``@away'' to higher charging availability at loading/unloading stops and during mandatory driver's breaks.}
    \label{fig_si22}
\end{figure}

In additional sensitivity analyses, we investigate to which extent BEV and ERS-BEV results are sensitive to alternative assumptions on charging availability. Our standard assumption is a grid connection of 200~kW at depots for both BEV and ERS-BEV. Pure battery electric trucks are assumed to also have charging options with power ratings of 200~kW at loading/unloading stops, and 500~kW during mandatory driver's breaks. While our depot charging availability assumptions may be overly optimistic (compare~\cite{speth_2024}), assumed fast-charging power ratings may conversely be considered too pessimistic, given recent and foreseeable advances in megawatt charging technologies (compare~\cite{shoman2023}). In sensitivities, we thus assume lower depot charging availabilities of 50\% or 25\% compared to our default parameterization (i.e.,~100 or 50~kW, labelled ``@depot''). Likewise, we assume higher charging availability of 200\% or 400\% compared to the default setting at loading/unloading stops (400 or 800~kW), and during mandatory driver's breaks (1000 or 2000~kW, labelled ``@away''). We illustrate the effects on power sector costs and electricity demand~(Figure~\ref{fig_si20}), the capacity mix and yearly electricity generation~(Figure~\ref{fig_si21}), as well as on optimal storage energy and power capacity~(Figure~\ref{fig_si22}). All figures show differences relative to a reference case without electrified HDV.

We find that a lower depot charging availability hardly has any effects on power sector costs, the optimal capacity mix, and yearly dispatch for the case of ERS-BEV, independent of the charging strategy. The same is true for inflexibly charged battery-electric trucks (BEV Inflex). The reason for this is that depot charging is hardly used at full capacity in any of these cases under default assumption. In the case of flexibly charging battery-electric HDV without vehicle-to-grid (BEV flex), power sector costs slightly increase with lower depot charging availability, as shifting grid consumption to periods with the lowest prices becomes less feasible. In the case with vehicle-to-grid (BEV Flex V2G), this cost increase is more pronounced, as the potential for using BEV batteries as distributed grid storage for solar PV integration decreases. Accordingly, optimal solar PV capacities also decrease with lower depot charging availability. Effects on optimal electricity storage capacities remain very limited in all cases.

Conversely assuming a higher charging power rating at loading/unloading stops and during driver's breaks hardly has any effects on power sector outcomes. We find a slight decrease in electricity sector costs, and very limited effects on optimal generation and storage capacities. This is because charging at these locations takes place in a very limited number of hours, and thus matters less from a power sector perspective as compared to charging at depots.

Overall, these sensitivity analyses suggest that alternative assumptions on charging availability generally have limited effects on results, which highlights the robustness of our main findings. Among all cases considered, vehicle-to-grid operations of flexibly operated BEV appear to be most affected by restricted depot charging options.

\clearpage

\putbib
\end{bibunit}

\end{document}